Dynamics of incoherent exciton formation in Cu$_2$O: Time- and angle-resolved photoemission spectroscopy

H. Tanimura[1,*], K. Tanimura[1,2,#], and P. H. M. van Loosdrecht[3]

1) Research Center of Ultra-High Voltage Electron Microscopy, Osaka University, Mihogaoka 7-1, Ibaraki, Osaka 567-0047, Japan
2) The Institute of Scientific and Industrial Research, Osaka University, Mihogaoka 8-1, Ibaraki, Osaka 567-0047, Japan
3) Physics Institute 2, Faculty of Mathematics and Natural Sciences, University of Cologne, Köln, 50937, Germany

We study dynamics of incoherent exciton formation under interband photoexcitation in Cu$_2$O using time-and angle-resolved photoemission spectroscopy at 90 K. Hot electrons injected by allowed optical transitions with 3.40-eV photons show ultrafast relaxation to the conduction-band minimum (CBM), surviving up to 500 fs after excitation. While hot-electron states with high excess energy show a rapid population decay of ~25 fs, an abrupt increase to 130 fs is observed for states with excess energies of 0.15 eV. This latter is interpreted in terms of phonon bottleneck dynamics characteristic of LO-phonon mediated energy relaxation. Excitons, having a small binding energy of 60 meV, are formed below the CBM quasi-instantaneously and subsequently relax to the 1S-exciton state of the yellow series within 1.5 ps. We find that, together with possible plasma screening for electron-hole interaction, the cooling of exciton center-of-mass motion plays an important role in the exciton relaxation dynamics. The characteristic features of exciton photoionization process are critically discussed based on the detailed analysis of angle-resolved photoemission results for the 1S excitons formed 1.5 ps after excitation.

**I. INTRODUCTION**

Energies, intensities and spectral line shapes in experimental optical spectra of semiconductors differ significantly from those predicted by the independent-quasiparticle picture (IQP) [1-3]. As is well known, excitonic effects provide the key concepts to gain a deeper understanding of the optical properties of solids, thereby providing the foundation for a wide range of optoelectronic and photovoltaic applications. The excitonic states play also crucial roles in relaxation processes of photoexcited semiconductors [3-5]. Under resonance excitation to excitonic optical transitions, ultrafast dynamics of dephasing and scattering processes of excitons in semiconductors have been studied extensively, and detailed knowledge of exciton dynamics in the coherent regime has been obtained [4, 5]. Under photoexcitation at higher energies which are off-resonant to excitonic transitions, oppositely charged electron and hole quasiparticles are generated in an ionized but correlated plasma, and they are combined into the excitonic states via dynamical relaxation processes in energy and momentum space [1, 6, 7]. Although the interplay of excitons and unbound electron-hole (e-h) pairs is at the heart of excited-state semiconductor physics, the ultrafast dynamics of the interplay remain elusive in many semiconductors.

For semiconductor quantum-well nano-layers, the dynamical interplay of optically-generated unbound e-h pairs and excitons has been studied directly using ultrafast terahertz spectroscopy which overcomes some limitations of the time-resolved optical techniques traditionally used [6]. Under non-resonant excitation above the band gap, an unexpected quasi-instantaneous excitonic enhancement and a slow exciton formation process over a 100-ps timescale has been revealed. Recently, similar ultrafast exciton formation and coexistence of excitonic states and e-h plasma under non-resonant excitation have been captured in single monolayers of $WSe_2$ [7]. How the characteristics of the interplay depend on the dimensionality of system, the electron-phonon interaction, the electron-electron interaction, and the e-h correlations is still an important issue to be explored. In this study, we employ time- and angle-resolved photoemission spectroscopy to investigate the relaxation dynamics of optically induced e-h plasma and the subsequent exciton formation processes in $Cu_2O$, where the lowest excitonic state has a binding energy as large as 150 meV [8, 9].

Using time-and angle-resolved photoemission spectroscopy, transient changes of photoinjected hot-electron populations in energy and momentum spaces have been studied with fs-temporal resolution, providing a direct view on hot electron relaxation processes [10-12]. The method has been applied to several semiconductors, and ultrafast relaxation processes of hot carriers in bulk electronic states have been revealed directly [13-23]. Because of the renewed interest in $Cu_2O$ with respect to solar-cell applications and as a promising p-type transparent conductive oxide [24], understanding of the hot carrier dynamics in this material is of great importance as an indispensable basis to improve efficiencies in optoelectronic device applications. In addition to this, photoemission spectroscopy for excitonic states provides a unique capability to obtain a detailed insight into the wavefunctions of excitons [25-27].

Recent experiments showing that the yellow-series excitons could be followed up to a high principal quantum number of n = 25 [28] boosted the renewed interest in excitons in $Cu_2O$. Observations like this, together with the ongoing research related to excitonic Bose-Einstein condensation [29], sparked extensive theoretical and experimental investigations on excitons in $Cu_2O$ in order to gain a deeper insight into spectroscopic features beyond a simple hydrogen-like model [30-34]. In these studies, the precise knowledge of wavefunctions of excitons, or the IQP Bloch states of which the excitonic state is composed, is of great importance. However, the correlation between an excitonic state and the IQP Bloch states responsible for the excitonic state, which links directly the excitonic picture and IQP band structures, has so far not been obtained experimentally. Time- and angle-resolved photoemission spectroscopy makes it possible, in principle, to determine these correlations experimentally [25-27].

In angle-resolved photoemission spectroscopy for electronic-structure determinations of solids, one tacitly assumes definite initial one-electron energy levels from which electrons are photoemitted. In the photoionization processes of excitons, however, such one-electron energy levels are indefinite because of the composite nature of excitons. Following the pioneering work on excitons by Elliott [35], an excitonic state $|n,\vec{K}\rangle$, characterized by the index n for the internal structure and the center-of-mass momentum wave vector $\vec{K}$, can be expanded in terms of IQP Bloch functions as [2, 35]

$$|n, \vec{K}\rangle = \sum_{\vec{k}_e, \vec{k}_h} f_{\vec{K}}^n(\vec{k}_e, \vec{k}_h) \hat{a}_{\vec{k}_e}^+ \hat{b}_{\vec{k}_h}^+ |0\rangle, \quad (1)$$

where $\hat{a}_{\vec{k}_e}^+$ and $\hat{b}_{\vec{k}_h}^+$ are creation operators of an electron with wave vector $\vec{k}_e$ in the conduction band (CB) and of a hole with wave vector $\vec{k}_h$ in the valence band (VB) of the IQP band structure, $|0\rangle$ is the vacuum state, and $f_{\vec{K}}^n(\vec{k}_e, \vec{k}_h)$ is the amplitude of constituent IQP band states. The IQP states involved are restricted by the relation $\vec{k}_e + \vec{k}_h = \vec{K}$. Therefore, for the exciton with $\vec{K} = 0$, $\vec{k}_h = -\vec{k}_e$. In the photoionization of excitons, the electrons are photoionized, but the conjugate hole states must restore their "original" valence-band states, inducing recoil effects on the photoemitted electrons, as depicted in Fig.1 for an exciton with $\vec{K} = 0$. Under energy and momentum conservation for the recoil energy $E_r(-\vec{k}_e)$ of the hole with momentum $-\hbar\vec{k}_e$, the photoemitted electron from the excitonic state has momentum $\hbar\vec{k}_e$ and the kinetic energy $E_K$ given by

$$E_K = h\nu_{probe} + E_{EX}^n(0) - \Phi_I - E_r(-\vec{k}_e), \quad (2)$$

where $h\nu_{probe}$ is the probe photon energy, $E_{EX}^n(0)$ the energy of excitonic state specified by the index n and $\vec{K} = 0$ measured from the valence-band maximum (VBM), and $\Phi_I$ the ionization energy of the material. It has been proposed that this recoil on hole has a statistical distribution along the momentum-energy curve of the VB of IQP band structure [26, 27]. On the other hand, the intensity of photoemission specified by $\hbar\vec{k}_e$ is proportional to $\left|f_0^n(\vec{k}_e, -\vec{k}_e)\right|^2$ such that angle-resolved photoemission spectroscopy directly measures the amplitudes of momentum- and energy-resolved IQP Bloch functions of which excitonic states are composed.

Time- and angle-resolved photoemission spectroscopy is also useful to capture dark-exciton dynamics. In optical spectroscopy for excitons, because of the polariton effects, only the excitons near $\vec{K} \approx 0$ are detected. On the other hand, photoemission spectroscopy can detect dark excitons with a finite $\vec{K}$ using the angle-resolved capability. The method was applied, among other, to study dynamics of excitons in organic and inorganic solids [36-40].

**II. EXPERIMENTAL**

The $Cu_2O$ with (111) crystal surface was a natural crystal, purchased from SurfaceNet GmbH. The (1x1) terminated surface was obtained by $Ar^+$-ion sputtering and annealing at 850-950 K in $O_2$ ($2 \cdot 10^{-6}$ mbar) [41, 42], and the surface structures were characterized in situ by a scanning tunneling microscope (STM). A typical STM image is shown in Fig.2. The surface-atomic images show a well-ordered (1x1) structure, together with randomly distributed protrusions with a typical size of ≈ 1 nm. The protrusions, denoted as P1 protrusions in Ref.41, were ascribed to surface oxygen vacancies [41]. The total number of surface defects including both the P1 protrusion and local dark spots was typically less than 10% of the surface sites surveyed. The samples were mounted on the six-axis manipulator of our photoemission instrument, and precisely oriented to the crystallographic directions determined by the STM image.

In order to gain bulk sensitivity in photoemission spectroscopy, one route is to increase the photon energy to the hard x-ray range [43], which raises high demands in obtaining sufficient energy and angular resolution. The alternative route is to use probe light at photon energies less than 10 eV [44], where the inelastic mean free path increases strongly [45] and the parallel-momentum resolution is improved drastically. In the present time-resolved photoemission measurements, the final state energies generated by pump and probe pulses were limited to below 8 eV with respect to the Fermi level, where the mean free path is typically 30 Å [45], which is substantially larger than the depth of the (1x1) surface layer of a few Å. A 76 MHz Ti-sapphire laser was used to generate fs-laser pulses at 730 nm, and the second and third harmonics of the fundamental were used as pump and probe beams in the time-resolved photoemission measurements, respectively. Their cross-correlation trace (CCT) in a BBO crystal had a full width at half maximum (FWHM) of 130 fs. The laser system was also used to generate the fourth harmonic of the 827 nm fundamental, which was used to determine the work function and the ionization energy of samples. Another laser system consisting of a Ti-sapphire laser oscillator, a regenerative amplifier, and a tunable optical parametric amplifier, generated 70-fs laser pulses centered at photon energy of 2.50 eV. A part of the amplified fundamental output at 795 nm was used to generate the third harmonic with a temporal width of 80 fs for probing photoemission. Pump and probe pulses, with a preset time delay (Δt), were aligned co-axially and focused on the sample surfaces at 45° to normal. The density $N_{ex}$ of excited states generated by a pump pulse was estimated using the formula; $N_{ex} = (1 - R)\alpha \Phi_{pump}$, where R, α, and $\Phi_{pump}$ are reflectivity, absorption coefficient, and the photon fluence at the photon energy of pump light. Based on the spectroscopic results of the complex dielectric constants [46], the pump-laser fluence was set to give excited-state densities typically $3 \times 10^{17}$ cm$^{-3}$, both for 2.50-eV and 3.40-eV photons. A hemispherical electron analyzer operating in an angle-resolved lens mode, and a two-dimensional image-type detector served as the electron spectrometer. Two-dimensional photoelectron images were recorded as a function of kinetic energy and emission angle of photoelectrons for $\pm 15°$ from surface normal. The energy resolution with fs-probe light was 80 meV, while the angular resolution was $\pm 1°$.

## III. RESULTS AND DISCUSSION

### A. General features of photoemission from hot electrons and excitons

Figure 3 (a) and 3(b) show the two-dimensional map of photoemission measured at Δt=17 fs and 500 fs after excitation with a pump-photon energy ($h\upsilon_{pump}$) of 3.40-eV. The probe-photon energy ($h\upsilon_{probe}$) is 5.10 eV. Photoemission intensities are plotted, with a color scale, as a function of energy ε and electron momentum ($k_\parallel$) parallel to the surface. The scale of ε is referenced to the VBM; ε was evaluated from the measured kinetic energy $E_K$ based on the relation $\varepsilon = \Phi_I + E_K - h\upsilon_{probe}$ as described in the Appendix A. Figure 3(c) shows a part of the one-electron band diagram of Cu$_2$O calculated by spin density functional theory [49]. The theoretical results have revealed unique features of non-parabolicity close to the Γ point in the top-most valence band (VB1), and have been used extensively to analyze theoretically the exciton structures in Cu$_2$O [31, 32]. However, the theory does not give the correct band-gap energy [47]. In Fig. 3(c), the band gap was set to be 2.17 eV for the calculated conduction-band dispersion, and the minimum of the second CB was set to be 2.62 eV, both based on experimental results [24]. We focus our attention only the dispersion along the Γ-X and the Γ-R directions because of the present experimental geometry described below.

Figure 3(d) shows the experimental geometry of the photoemission measurements, and the relation between the surface Brillouin zone (SBZ) and bulk Brillouin zone (BBZ) for (111) oriented $Cu_2O$ [47]. The [111] crystal axis is aligned along the surface normal, and the [11$\bar{2}$] and [111] directions define the detection plane. In the photoemission process, the $k_\parallel$ parallel to the surface, given by $k_\parallel = \sqrt{(2m_0/\hbar^2)E_K}\sin(\theta)$, is conserved ($m_0$ and $\hbar$ being the electron rest mass and Planck's constant) [48]. Therefore, the measured photoemission image shows a one-dimensional cut along $\bar{M}$-$\bar{\Gamma}$-$\bar{M}'$ of the SBZ which represents the two-dimensional projection of the three-dimensional electron distribution in the BBZ.

The projection has the following characteristics. First, all states along the Γ-R ([111]) direction in the BBZ are projected at $\bar{\Gamma}$, contributing to surface-normal photoemission. Second, all states along the Γ-X ([001]) direction contribute to off-normal emission with $k_\parallel = k_X \sin(54.75°)$ on the right half along the $\bar{\Gamma}$-$\bar{M}'$ direction (θ>0). Third, on the left half (θ<0) along the $\bar{\Gamma}$-$\bar{M}$ direction, the states along the Γ-M ([110]) in the BBZ are projected on the $\bar{\Gamma}$-$\bar{M}$ direction with $k_\parallel = k_M \sin(35.26°)$. The states along the Γ-R ([11$\bar{1}$]) direction contribute to the off-normal emission with $k_\parallel = k_R \sin(70.52°)$. Here $k_X$, $k_M$ and $k_R$ are wave vectors along the Γ-X, Γ-M and Γ-R directions in the BBZ. We focus mainly on the dispersive features measured at θ>0, as those at θ<0 include contributions of many states with different symmetries and dispersive characteristics, which make the analysis more complicated.

In Figs. 3(a) and 3(b), the dispersions of the CB along the Γ-X direction are plotted as a function of $k_\parallel$. The dispersion curve of the first CB roughly represents the lower boundary of the CB; all CB states lie above the curve. The pump light with photon energy of 3.40 eV induces allowed optical transitions from the VB to the second CB. Based on the band structure shown in Fig. 3(c), the highest-energy states reached by optical transitions can be predicted as shown by violet arrows; the highest energy state along the Γ-X direction is 3.3 eV, while that along the Γ-R direction is 3.2 eV. In the image in Fig. 3 (a), the hot electron population at Δt=17 fs shows a broad distribution extending from 2.6 to 3.3 eV; the energy range corresponds to the second CB in $Cu_2O$ into which hot electrons are photo-injected by the 3.40-eV light pulses. The initial states responsible for the photoemission are mostly distributed at ε>2.17 eV, although a small fraction below the CBM can be detected even at Δt=17 fs. On the other hand, as seen in Fig. 3(b), the one intense peak below the CBM (called EX photoemission hereafter) dominates over others at Δt=500 fs, showing ultrafast changes in photoemission characteristics within the first 500 fs.

Figure 4 shows the temporal evolution of angle-integrated photoemission spectra; photoemission intensities are integrated with respect to emission angles from $0°$ to $+15°$. Figure 4(a) highlights the changes above the CBM on an expanded vertical scale, while Fig. 4(b) displays the changes in spectra below 2.2 eV. In Fig. 4(a), it is evident that the intensities from states above the CBM are lost within ~500 fs. The loss in the intensity from the hot electrons is accompanied by a strong enhancement of the photoemission peak below the CBM. Importantly, the EX photoemission with a higher peak energy is formed quasi-instantaneously even at Δt=17 fs. The intensity of EX peak increases dramatically at larger Δt's, and the growth of the EX photoemission peak is associated with significant red shift of the peak energy $E_P$: $E_P$=2.10 eV at Δt=17 fs, while it is 2.04 eV at Δt=1.5 ps. The $E_P$ and the intensity of EX photoemission become unchanged at Δt=1.5 ps, showing the formation of a quasi-steady state. The peak energy of 2.04 eV coincides well, within the experimental energy resolution, with the 1S exciton state of the yellow exciton series (the distinction between ortho- and para-exciton states is not possible under the present energy resolution). Based on these observations, we attribute the EX photoemission to photoionization of the excitons in $Cu_2O$.

The results in Figs. 3 and 4 show clearly that a e-h plasma formed by photoexcitation with 3.40-eV photons are transferred into excitonic states in the time frame of ~1.5 ps after excitation, and a quasi-steady state of excitons is formed at Δt=1.5 ps. Characteristic features of hot-electron decay and exciton-formation dynamics are the main topics in the latter sections. Prior to analyzing the results on the dynamics and discussing the consequences in detail, we first focus our attention to the characteristics of the excitonic photoemission in III-B, based on results at Δt=1.5 ps when a quasi-steady state is established. Then, based on a quantitative analysis of earlier time results we discuss the dynamics of hot-electron relaxation in the CB in III-C, and the incoherent exciton-formation dynamics in III-D.

**III-B. Characteristics of exciton photoionization**

As described in Eqs. (1) and (2), photoionization of an exciton with $\vec{K} \approx 0$ generates a photoelectron with a kinetic energy that is governed by recoil effects of conjugated holes via $E_r(-\vec{k}_e)$. The momentum-resolved (k-resolved) intensity of photoemission is determined by $\left|f_0^n(\vec{k}_e, -\vec{k}_e)\right|^2$, which is the squared amplitude of exciton wavefunction. Therefore, the k-resolved kinetic energy and intensity of photoemission are unique and important quantities obtained by photoemission spectroscopy for excitons.

(B・1) The dispersion of $E_K$ in exciton photoionization

Figure 5(a) displays the two-dimensional map of photoemission measured at Δt=1.5 ps; the photoemission intensities specified by a color scale are plotted as a function of $E_K$ and $k_\parallel$. In the photoionization of excitons with $\vec{K} = 0$,

$$k_\parallel = \sqrt{\frac{2m_0}{\hbar^2}\{h\upsilon_{probe} + E_{EX}^n(0) - \Phi_I - E_r(-k)\}} \sin(\theta). \tag{3}$$

The determination of $k_\parallel$ fixes a point on the two-dimensional SBZ; $k_\perp$ can have a value anywhere along the rod extending into the three-dimensional BBZ [48]. In the present case for θ>0, we can specify the directions from the Γ to any point I on a zone edge in the BBZ by introducing the angle $\gamma_I$ ($0° < \gamma_I < 90°$), which is the angle between the $\Gamma$-R[111] direction and $\Gamma$-I direction inside the plane defined by [111] and [11$\bar{2}$] directions (see Fig.3(d)). Using the wave vector $k_I$ along the $\Gamma$-I direction, $k_\parallel = k_I \sin(\gamma_I)$. Therefore, the off-normal photoemission detected experimentally at a given $k_\parallel$ is a superposition of many components from such states that are projected on the one-dimensional cut along $\bar{\Gamma}$-$\bar{M}'$ direction of the SBZ.

Despite the composite features in off-normal emission mentioned above, it is clear that the kinetic energy decreases with increasing $k_\parallel$, although the effect is small. The dispersion of $E_K$ in Fig. 5(a) reflects the dispersion of $E_r(-\vec{k}_e)$, which is determined by the dispersion of the VB (see Fig. 1) [26, 27]. It is instructive to compare the observed dispersion of $E_K$ with the theoretical dispersion of the VB1 band. As seen in Fig. 3(d), the states along the $\Gamma$-X direction contribute to the off-normal emission with $\gamma_I = 54.75°$, and the dispersion of VB1 has been calculated along the $\Gamma$-X direction in Ref. 47. Using the calculated valence band dispersion, $E_K$ as a function of $k_X$ can be determined from Eq. (2) for $E_r(-k_X) = E_{VB1}(0) - E_{VB1}(-k_X)$. Combined with the relation $k_\parallel = k_X \sin(54.75°)$, thus determined $E_r(-k_X)$ can be plotted as a function $k_\parallel$. The solid curve in Fig. 4(a) shows the calculated dispersion, which coincides reasonably well the experimentally observed small downward dispersion of the kinetic energy.

The contributions to off-normal emission at a given $k_\parallel$ from the states along the directions with $\gamma_I < 54.75°$ are expected to show larger amounts of dispersion, as the magnitudes of $k_I$ become larger. However, because of the k-dependent photoemission intensities, which rapidly decrease with increasing $k_I$ as discussed in the next paragraph, the highly dispersive components may not be detected easily. On the other hand, the contributions from the states along the directions with $\gamma_I > 54.75°$ may result in less dispersive photoemission. However, because of a finite energy resolution of 80 meV in the present study, it is not possible to resolve such components with different dispersive characteristics. Therefore, we can conclude that the dispersion of photoemitted electrons from the 1S excitons is characterized by the downward dispersion, reflecting recoil effects of the conjugated holes, and that the theoretical results of spin-density functional theory of the top-most valence band dispersion is representative for the dispersion characteristics.

(B・2) The k-dependent photoemission intensity in exciton photoionization

In Fig. 5(b), the energy-integrated (1.2 eV< $E_K$ <1.45 eV) photoemission intensity from the 1S excitons is plotted as a function of $k_\parallel$; the intensity decreases with increasing $k_\parallel$. In order to gain insight into the momentum-resolved characteristics of exciton photoemission, we analyze the result in terms of a hydrogenlike 1S state $\Psi_{1S}(r)$ for the exciton. Even though the exciton ground state will show deviations from $\Psi_{1S}(r)$, it well describes the radial probability density [32]. For $\Psi_{1S}(r)$, the modulus squared, $|\Phi_{1S}(\bar{k})|^2$, of the Fourier transform of $\Psi_{1S}(r)$ is given by

$$|\Phi_{1S}(\bar{k})|^2 = \frac{8(a_{ex})^3}{\pi^2\{1+k^2(a_{ex})^2\}^4} \tag{4}$$

which corresponds to $|f_0^{1S}(\vec{k}_e, \vec{k}_h)|^2$ in Eq. (2). In Eq. (4), $a_{ex}$ is the Bohr radius of the 1S exciton state [9, 32], and $k$ is the wave vector from the Γ point in the BBZ. Because of the symmetry of $\Psi_{1S}(r)$, the wave vector $k_I$ along the Γ–I direction in the BBZ satisfies Eq.(4). Therefore, the intensity in Fig. 4(b) can be described as a superposition of $|\Phi_{1S}(k_I)|^2$ for different $k_I$'s with the same $k_\parallel$. When we assume that each component along any direction contributes to the photoemission intensity with an equal weight, the experimentally determined photoemission intensity $I(k_\parallel)$ as a function of $k_\parallel$ is given by

$$I(k_{//}) = \int_0^{90} \frac{8(a_{ex})^3}{\pi^2[1+\{k_\parallel/\sin(\gamma_I)\}^2(a_{ex})^2]^4} d\gamma_I . \tag{5}$$

The magnitude of $a_{ex}$ reported in recent literature ranges from 5.3 Å [9] to 8.0±0.1 Å [32]. The solid black curve shows the result of Eq. (5) obtained by numerical integration for $a_{ex}$=5.3 Å.

Similar results for other magnitudes of $a_{ex}$ are also shown for comparison. For $a_{ex}$=6.5 Å, the calculated intensity is too small for entire region of $k_{\parallel}$, while for $a_{ex}$=4.5 Å, the calculated intensity is much larger than the experimental result at large $k_{\parallel}$ region. Therefore, a hydrogenlike 1S-state wavefunction with $a_{ex}$=5.3 Å gives a reasonable basis on which k-resolved photoemission intensities can be described.

To support this picture without having to assume that all states contribute equally to the photoemission, we also analyzed the normal photoemission intensity at Δt=1.5 ps using the same hydrogenlike 1S-state wavefunction. As shown in Fig. 3(d), the surface-normal photoemission is contributed from the states along the Γ-R direction only, allowing us to directly relate the kinetic energy and $k_R$ using the band-structure calculations. As defined in Eq. (2), the kinetic energy of the photoelectrons in the exciton photoionization is governed by the recoil energy of the conjugated holes. In the present case, the recoil energy is determined by the dispersion of the top-most valence band along the Γ-R direction; $E_r(-\vec{k}_e) = E_r(-k_R)$. This dispersion is displayed in Fig. 6(a). The photoemission intensity with a given $E_r(k_R)$ is determined by $|f_0^{1S}(k_R,-k_R)|^2$, which is given by Eq.(4) for the hydrogenlike 1S-state wavefunction. The theoretical spectrum $|f_0^{1S}(k_R,-k_R)|^2$, using $a_{ex}$=5.3 Å, is shown by solid red curve in Fig. 5(b). The solid green curve in Fig. 6(b) shows the surface-normal photoemission spectrum at Δt=1.5 ps. The peak is clearly asymmetric; the low-energy side of the peak is wider than the high-energy side. The solid black curve, which shows the calculated spectrum convoluted with respect to the finite energy resolution of 80 meV, reproduces reasonably well the experimental normal photoemission spectrum, thereby confirming that the characteristic features of the photoemission from the 1S excitons in $Cu_2O$ are consistently described by the theoretical framework of exciton photoionization as discussed in Refs.26 and 27 using a hydrogenlike 1S-state wavefunction with $a_{ex}$=5.3 Å.

**C. Relaxation dynamics of hot electrons in the conduction band**

In order to capture characteristic features of hot-electron relaxation in the CB, we first focus our attention to temporal changes in populations at different energy- and momentum-resolved states. In Fig. 3(a), we introduce seven regions labeled A to G. As the regions B, E, and G monitor the surface-normal emission, temporal changes represent changes in population of states along the Γ-R direction at different energies. On the other hand, intensities at the regions A, C, D and F represent temporal changes in hot-electron populations at the states with finite values of $k_{\parallel}$. For the states along the Γ-X direction, the regions A and D corresponds the states within the CB2, while the regions C and F are located within the CB1. In order to specify the energy of hot-electron states, we use the excess energy $E_{ex}$ referenced to the CBM (2.17 eV above the VBM). For the Regions A and B $E_{ex}$ =0.96 eV, for the regions A, D, and E $E_{ex}$ =0.63 eV, and for the regions F and G $E_{ex}$ =0.26 eV.

The temporal changes in photoemission at the regions introduced above are plotted as a function of time delay in Figs. 7(a), 7(b) and 7(c). It is clear from these figures dynamics of the hot electron populations are depend on $E_{ex}$, but are, for a given $E_{ex}$, independent of the wave vectors in the BBZ. These observations strongly suggest that the hot electron populations are quasi-equilibrated in the momentum space within a pump-pulse temporal width, while they are still non-thermal in the energy space. Such ultrafast momentum relaxation as observed here has also been reported for GaAs [50, 51], Si [52], and GaN [53] and is thought to be due to ultrafast momentum scattering by electron-phonon interaction leading to the formation of a so called hot-electron ensemble [50]. Also for oxide semiconductors like ZnO [54] and TiO$_2$ [55], first-principle theoretical studies have reported ultrafast momentum relaxation times of a few fs due to electron-phonon interaction. In addition to the relaxation mediated by the electron-phonon interaction, it is clear that also electron-electron and electron-hole interactions among hot carriers is expected to contribute to establishing quasi-equilibration in the momentum space on an ultrafast time scale.

Figures 7(a)-(c) show that the population relaxation speeds up for states with higher energy $E_{ex}$. We determine the decay time of the states with a given $E_{ex}$ by analyzing the angle-integrated

($0°$-$15°$) transient intensities at various $E_{ex}$ (integrated over $\pm 40$ meV); the integration does not change any physics because of the k-independent dynamics at a given $E_{ex}$. The angle-integrated transient intensities at various $E_{ex}$ are plotted in Fig. 7(d) on a semi-logarithmic scale. The broken curve in the figure is the Gaussian-fitted CCT between pump and probe pulses with the width of 130 fs. As the intensity of CCT falls down to less than 1/100 of the maximum intensity at Δt=120 fs, we can determine the decay time of the hot-electron population at a given $E_{ex}$ using the empirical semi-logarithmic plot of the intensities at Δt >120 fs. This method can be applied to the data of $E_{ex}$ <1.0 eV, which show high enough intensities at Δt >120 fs for the quantitative analysis. As the photoemission intensities at $E_{ex}$ >1.0 eV decay too fast to apply this empirical method; we used a rate-equation analysis to estimate the decay time. The solid line in the figure is a fit to the data at $E_{ex}$ =0.13 eV, giving a decay time of 87 fs. An example of the rate-equation analysis is shown by solid black curve for the data at $E_{ex}$ =1.15 eV.

The resulting population decay times are plotted as a function of $E_{ex}$ in Fig. 8(a); Fig. 8(b) portrays the dispersions of conduction band along the Γ-X direction. The decay time, which is 26 fs at $E_{ex}$ =1.15 eV, becomes gradually longer with decreasing $E_{ex}$, and it increases quite abruptly at $E_{ex}$ <0.2 eV. As seen in the temporal changes in photoemission spectra in Fig. 4(a), the hot-electron population above the CBM decays within a few hundreds of fs. Such extremely fast electron relaxation of hot electrons above the CBM has also been detected experimentally in other transition metal oxide crystals like ZnO [38] and TiO$_2$ [56]..

The above determined population decay times represent the energy relaxation of the photoexcited state (through $E_{ex}n_{ex}$, with n$_{ex}$ the population at $E_{ex}$). Describing energy relaxation processes of hot electrons in the electronic system with strong electron-hole correlations is not trivial. However, it is instructive to discuss the energy relaxation process in terms of expectations from a Fröhlich electron-LO phonon interaction mechanism based on the IQP picture, as in Refs. 54 and 55.

It has been reported that the density of states of the highest LO-phonon branch in Cu$_2$O shows a peak at 74 meV [57]. We use this energy to evaluate the energy relaxation rate by the Fröhlich interaction. As the LO-phonon energy is substantially higher than the thermal energy of 90K (7.5 meV), stimulated phonon processes can be ignored. In such a case, taking into account non-parabolicity in the CB, the energy relaxation rate for the electron energy $e_k$ (specified by the wave vector k) referenced to the CBM, is given by [58]

$$\frac{de_k}{dt} = -\frac{e^2(2m^*)^{1/2}\omega_{LO}^2}{4\pi\varepsilon_p \gamma^{1/2}(e_k)} I^2(k,k') \left[ \left\{\frac{d\gamma(e_k)}{de_k}\right\}_{e_k - \hbar\omega_{LO}} \tanh^{-1}\left\{\frac{\gamma^{1/2}(e_k - \hbar\omega_{LO})}{\gamma^{1/2}(e_k)}\right\} \right], \quad (6)$$

where $m^*$ the effective mass of the CB, $\omega_{LO}$ the LO-phonon frequency, $I(k,k')$ the overlap integral of Bloch functions of initial and final states, $1/\varepsilon_p = 1/\varepsilon_\infty - 1/\varepsilon_s$ with the high frequency $\varepsilon_\infty$ and static $\varepsilon_s$ dielectric constants, and $\gamma(e_k)$ is defined as

$$\gamma(e_k) = \hbar^2 k^2 / 2m^* = e_k(1 + \alpha e_k) \quad (7)$$

with a constant $\alpha$ that characterizes non-parabolicity. We analyzed the theoretical result of the first CB of Ref.49 to determine $m^*$ and $\alpha$ using Eq. (7). For the energy range from 0 to 0.6 eV above the CBM, the dispersions along Γ-X, Γ-M, and Γ-R directions can be well characterized by $\alpha =$ 0.14±0.03 (eV)$^{-1}$ and $m^*=(0.93\pm0.01)m_0$. Although $I(k,k')$ is less than unity where non-parabolicity is present, we assumed $I(k,k')=1$ for simplicity.

Using these results, we determined the population decay time as $\tau = \Delta E/(de_k/dt)$ with $\Delta E$ =80 meV as also used in the experimental population-decay analysis. Like the experimental and in line with the expected "phonon bottleneck" when approaching the CBM [54, 55], the resulting theoretical curve, plotted in Fig. 8(b) (solid line), shows an abrupt slowing down at low energy, while the rate is almost constant above $E_{ex}$~0.1 eV. In contrast to the theoretical curve, the experimentally determined population decay time becomes shorter with increasing $E_{ex}$ above ~0.4 eV. This increase in the energy relaxation rate can be attributed to the increase of final density of states in the LO-phonon scattering in the phase space by the presence of second and third conduction bands above 0.45 eV; Eq. (7) does not include any effects caused by these higher bands.

Though it is clear that Fröhlich electron-LO phonon interaction describes the main features of the relaxation qualitatively, it is also clear that this interaction is not solely responsible for the hot electron relaxation as the electrons in the CB are lost by combining with holes to generate excitons. In order to have some insight into the dynamics of e-h combination to form excitons, we calculated the energy- and angle-integrated photoemission intensity [59]. This total photoemission intensity, $I_{total}$, plotted in Fig. 9(a) (green line), is lost within 500 fs after excitation. Using a rate-equation analysis, we find that the temporal change of $I_{total}$ can be described approximately by an effective decay time of 130 fs. On the other hand, the total density $N_e$ of hot electrons generated by a pump pulse can be estimated by the integration of the light-pulse shape with the same parameters in the rate-equation analysis of $I_{total}$. The estimated $N_e$ is shown by the broken curve in Fig. 9(a). The difference between $N_e$ and $I_{total}$ is representative of the temporal change in the loss of electron density from the CB. Assuming that electron-hole recombination leading to their full annihilation does not occur on these short time scales, it is expected that the total density of excitons with any internal structures with any $\vec{K}$ may follow such growth kinetics shown by the broken curve, which reaches a constant value around $\Delta t=500$ fs (the solid red curve).

**D. Dynamics of incoherent process of exciton formation**

Based on the analysis of hot-electron relaxation described in the above section, it is estimated that the e-h combination to form excitons occurs within 500 fs of excitation. On the other hand, as shown in Fig.4, EX photoemission exhibits gradual changes not only in intensity but in the peak energy in the time frame of 1.5 ps. In Fig. 9(a), the $I_{total}$ of the EX photoemission [59] is plotted as a function of $\Delta t$; it increases to a maximum around $\Delta t=1.5$ ps [60]. In Fig. 9(b), the $E_P$ of the normal photoemission spectra ($\theta=\pm 1°$) is plotted as a function of $\Delta t$; the normal photoemission spectra are chosen in this analysis in order to exclude any contributions of dispersive components with finite $k_{\parallel}$. The $E_P$ shifts from 2.10 eV at $\Delta t=120$ fs to a constant value of 2.04 eV at $\Delta t=1.5$ ps. The results reveal that the relaxation process of excitons leading to 1S-exciton states take place in the time frame of 1.5 ps after e-h combination within 500 fs.

As the $E_P$ of photoemission reflects the binding energy of an exciton, the results in Figs. 4 and 9 show that the wavefunctions of excitons are changing toward that characteristic of the 1S state in the time frame of 1.5 ps. One possible scenario of the excitonic-structural changes could be a relaxation cascade within the excitonic manifolds from high Rydberg states to the lowest state, as proposed for MoSe$_2$, in which the lowest exciton has the binding energy similar to that in Cu$_2$O [61]. In fact, the energies of 2p- and 2s-exciton states are located within the gap, as shown in Fig. 9(b) [62]. However, the $E_P$ of EX photoemission at Δt=17 fs is 2.10 eV, lower than the energies of 2s- and 2p-exciton states, and the $E_P$ lies between 2.10 to 2.04 eV during the relaxation. One may argue that the energy resolution in the present measurements is not good enough to resolve the separation of 1S and 2S exciton peaks of 130 meV; a single broad peak, instead of two separated peaks corresponding to the photoemissions from 1S and 2S (2p) states, could be observed because of the finite energy resolution. In order to examine such possibilities due to a finite energy resolution, we performed a deconvolution analysis of the observed photoemission spectra with respect to the energy resolution of 80 meV (see Appendix B). The results of deconvolution clearly show a single photoemission peak; two-peak structures could not be resolved for any spectra at Δt ranging from 17 fs to 1.5 ps. Therefore, the results in Figs. 4 and 9 indicate that transient excitonic states with time-dependent wavefunctions are relaxing by taking different pathways from the relaxation cascade within the excitonic manifolds.

We show two-dimensional maps of EX photoemission at early time delays; photoemission intensities specified by a color scale are plotted as a function of $E_K$ and $k_{\parallel}$ in Figs. 10 (a), 10 (b), and 10 (c). Figs. 10(d), 10(e), and 10(f) show k-resolved photoemission spectra at $k_{\parallel}$=0 and 0.10 Å$^{-1}$ in the images in Figs. 10(a), 10(b) and 10(c). It is clear that the $E_K$ shows upward dispersion at Δt=120 fs and at 205 fs; the peak energy is higher at larger $k_{\parallel}$. At Δt=320 fs, the k-resolved photoelectron distribution of the EX photoemission is almost flat; the energy difference between $k_{\parallel}$

=0 and 0.10 Å$^{-1}$ is less than 0.01 eV. The $E_K$ finally shows the downward dispersion at Δt=1.5 ps, as shown in Fig.4.

As analyzed extensively in the theoretical study of exciton photoemission, the upward dispersion of $E_K$ as a function of $k_\parallel$ is a clear indication that excitons with finite K ($=|\vec{K}|$ of wave vectors of center-of-mass motion) are excited [27]. In the photoionization of an exciton with a finite K, the highest $E_K$ of photoelectron is expected at $\vec{k}_e = \vec{K}$, as the recoil energy of hole is zero ($\vec{k}_h = 0$), and the exciton energy becomes higher by $\hbar^2 K^2/2M$ (M is the mass of center-of-mass motion) than that of K=0 at $\vec{k}_e = \vec{K}$. The $E_K$ is associated with an asymmetric downward dispersion by the recoil effects of conjugated holes. However, when excitons with different K's are populated continuously, and when the amount of downward dispersion of each component is small as in the present case, momentum-resolved photoelectron distributions from an ensemble of excitons may look like as if photoelectrons lie on the exciton dispersion curve characterized by M and K. For $M = m_e^* + m_h^* = 1.57 m_0$ [24] and $K = k_\parallel$, we evaluated the exciton dispersion, and the results are shown by solid white curves in Figs. 10(a), 10(b) and 10(c). At $k_{//} = 0.10$ Å$^{-1}$, $\hbar^2 K^2/2M = 0.029$ eV, which is nearly the same amount we observe in Figs.10 (a) and 10(c).

Figure 10(g) shows the k-resolved intensity distribution of EX photoemission at Δt=120 fs, 205 fs, 320 fs, and at 1.5 ps; the intensity is normalized at the value at $k_\parallel$=0. The distribution is broader at early time delays, which can be interpreted reasonably as a superposition of contributions from excitons with many different $\vec{K}$. Relatively high intensities at large $k_\parallel$ at early time delays are indicative of efficient generation of excitons with larger K. Interestingly, the distribution at Δt=120 fs shows a peak not at $k_\parallel$=0 Å$^{-1}$ but at $k_\parallel$=0.063 Å$^{-1}$, suggesting non-thermal distribution of excitons with respect to K. The progressive narrowing of the distribution width shows the cooling of center-of-mass motion of excitons in the exciton band. The half width of the k-resolved intensity distribution in Fig. 10(g) is plotted as a function of Δt in Fig. 9(c). The width becomes narrower within the first 400 fs rapidly, with further persistent narrowing till Δt=1.5 ps. The rapid process may result from interactions of transient excitons with LO phonons, while the slow process may be due to interactions with acoustic phonons, which eventually leads to the excitonic order characteristic of the lowest 1S-exciton state in the time frame of 1.5 ps.

The high peak energy (or low binding energy) of EX photoemission at early time delays may reflect the transient nature of the correlations as mediated by Coulomb interactions; excitonic states may be formed from unbound e–h pairs transiently by transferring energy and momentum to other carriers [6]. At Δt<500 fs, such transiently formed excitons and an e-h plasma can coexist under 3.40-eV excitation. Therefore, possible screening in the electron-hole interaction may reduce the binding energy of excitons to form only excitonic states with large spatial extensions. However, this screening effect cannot explain the whole process of relaxation in the time frame of 1.5 ps. First, as seen in Fig. 9, changes in the peak energy and the EX-photoemission intensity last at Δt>0.5 ps, within which the electron population in the CB decays. Second, as shown in Appendix C, the peak energy and the total intensity of the EX photoemission under excitation at 2.50 eV, where the initial density of e-h plasma is substantially lower than the case of 3.40-eV excitation, show similar changes to those under 3.40 eV excitation. Therefore, the cooling of center-of-mass motions of excitons also plays an important role in enhancing exciton binding energy in the transient process lasting up to 1.5 ps. The exciton binding energy of 150 meV is a consequence of perfect excitonic order in $Cu_2O$. Therefore, any disturbances to the order, like a finite K, may reduce the amount of binding energy. Although this problem may have close correlation with an important issue of K-dependent wavefunction of excitons, we leave it as an open question before establishing more precise correlations between the magnitude of K and the $E_P$ of excitons in $Cu_2O$.

Finally, we discuss briefly the difference between the temporal evolution of the $I_{total}$ of EX photoemission and the predicted growth of exciton density shown in Fig.9. As described above, excitonic states formed by electron-hole combination initially have lower binding energy, which corresponds to the large spatial separation between electrons and holes in the relative space. The IQP

Bloch states participated in the exciton wavefunctions are the electron and hole states which are distributed in relatively narrow range in the momentum space, but are largely separated by a finite K. As the relaxation proceeds, the spatial separation becomes shorter with increasing binding energy. Then many IQP Bloch states near the center of BBZ become responsible for the excitonic states. Even for the transient process involving changes in exciton wavefunctions, an ideal sampling of all possible excitonic states over the whole BBZ will give a quantity which is proportional to the density of excitons because of the normalization of exciton wavefunctions. However, we could not carry out such an ideal sampling because of the limited crystal orientation and limited detection angles in the present study. Under the present experimental geometry shown in Fig. 3 (d), photoemission intensities were probed only for a two-dimensional plane (determined by the detection plane) within a three dimensional distribution in the BBZ. In this case, the fraction of excitons, probed experimentally as $I_{total}$, depends on the momentum spread of excitonic states in the k space; the fraction is roughly proportional to the inverse of momentum spread. Therefore, the measured $I_{total}$ of the EX photoemission may include such a k-dependent factor that can cause the difference between the measured $I_{total}$ and the exciton density. It is expected that future time-and angle-resolved photoemission spectroscopy studies with higher energy resolution for a wider momentum space can overcome the experimental limitations and reveal essential details of the exciton formation dynamics under band-gap excitation in semiconductors.

V. Summary

Using time- and angle-resolved photoemission spectroscopy, we have studied the dynamics of incoherent exciton-formation process under interband photoexcitation in $Cu_2O$. Hot electrons injected more than 1.0 eV above the CBM show ultrafast energy relaxation with a time constant typically a few tens of fs, similar to other semiconductor oxide crystals. Qualitative features of the hot-electron energy relaxation, including a phonon bottleneck, have been described in terms of the Fröhlich interaction. Formation of excitonic states shows a quasi-instantaneous enhancement, but the initially formed excitonic state is characterized by a smaller binding energy of 60 meV. The full transformation from the primitive exciton state to the lowest 1S excitonic state takes place in the time frame of 1.5 ps. Our study suggests that the exciton relaxation process is not the relaxation cascade in the exciton manifolds involving higher Rydberg states of 2s and 2p states as discussed for MoSe$_2$, but rather is governed by the cooling of the exciton center-of-mass motion, together with screening of the electron-hole interaction by the photoinduced e-h plasma. We have further shown that the photoemission results for the 1S excitons in Cu$_2$O are consistently described using the theoretical framework discussed in Refs.26 and 27. In particular, the momentum-resolved intensity of exciton photoionization gives direct access to the radial probability density of excitons, and the momentum-resolved kinetic energy of exciton photoionization provides the dispersive features of the valence-band state involved in exciton states. Thus time- and angle-resolved photoemission spectroscopy provides experimental results revealing momentum-resolved characteristics of exciton states. Though the energy- and angle-integrated intensity of exciton photoemission is shown to be very sensitive to the wavefunction of excitons, even more detailed time- and angle-resolved photoemission studies, together with advanced theoretical studies, are needed to allow for a full characterization of the exciton wavefunctions in energy and momentum space.

Acknowledgement

We thank K. Nasu, J. Kanasaki, and J. Güdde for valuable discussions. This work was supported by the Japan Society for the Promotion of Science (JSPS) KAKENHI Grant No. 24000006.

**Appendix A: The ionization energy of Cu$_2$O (111)-(1x1) surface**

In order to correlate photoemission peaks with any states in bulk electronic structures, the exact value of ionization energy $\Phi_I$ is crucial. A theoretical study, using density functional theory with a hybrid functional, has reported the value of 5.18 eV [63], which is close to the experimental value (5.21 eV) determined by photoelectric threshold measurements for single-crystalline Cu$_2$O with (111) surfaces [64]. However, as the magnitude of $\Phi_I$ depends critically on the potential associated with the electric dipole layer at the real surface, including surface reconstructions and degree of surface perfection [64], it is important to determine $\Phi_I$ for the sample used in the present study directly. Here we determined the magnitude of $\Phi_I$, by photoemission spectroscopy using 6.00-eV laser light for the samples used in the time-resolved study.

Figure 11 shows the photoemission spectrum probed by p-polarized 6.00-eV laser light for $Cu_2O$:(111)-(1x1) at 90 K. In order to establish surface conditions similar to those in time-resolved measurements, the sample was excited with 3.00-eV laser pulses at $\Delta t$= -5 ps. As the repetition rate of laser pulses is 76 MHz, the 3.00-eV laser pulses excite samples 13.1 ns before probing photoemission, which reduces possible surface band bending via photovoltaic effects. In the spectra shown in Fig.6, the electron kinetic energy $E_K$ is referenced with respect to the vacuum level of the electron analyzer. There is a clear low-energy cutoff $E_L$ at $E_K$ =0.830 eV, which corresponds to the difference between the vacuum level of $Cu_2O$:(111)-(1x1) and the vacuum level of the electron analyzer. As the work function of the analyzer is calibrated to be 4.337 eV, the spectrum shows that the work function of the sample is 5.17 eV. At the higher energy region ($> E_L$), the intensity decreases almost linearly to a constant level with a finite intensity. We determine the high-energy cutoff $E_H$ as the crossing point of a line at $E_K$ from 1.0 to 1.1 eV with the constant weak-intensity level, as shown in the figure. We identify the $E_H$ =1.05±0.01 eV to be the onset of photoemission from the VBM probed by 6.00-eV photons. The quasi-constant weak photoemission can be ascribed to the mid-gap states. Then, the Fermi level $E_F$ is determined to be 0.61±0.01 eV, giving the ionization energy of 5.78±0.01 eV.

Electronic states of semiconductors often shows band bending near the surfaces, and photoexcitation changes the amount of band bending via photovoltaic effects. However, as all states in the surface region monitored by 2PPE shift their energies in parallel [65], use of $E_K$ referenced by the $E_L$ gives a more reliable energy axis to analyze the spectra in terms of bulk electronic states by compensating for possible variations of the photovoltage originating from different excitation and/or probe conditions. Then for a photoemission peak with $E_K$, the energy ε of the initial IQP state of photoemission referenced with the VBM is determined as $\varepsilon = \Phi_I + E_K - h\upsilon_{probe}$.

**Appendix B: Deconvolution analysis of photoemission spectra**

In the present study of photoemission measurements, energy resolution $\Delta E$ was limited to 80 meV. Therefore, any structures in photoemission are detected as convolution of the true spectra with respect to $\Delta E$. In most cases, we analyzed and discussed the results by taking the finite energy resolution into account. However, in some cases, a deconvolution analysis becomes crucial to reveal specific features of photoemission. In the deconvolution analysis, we applied a simple method to fit an observed photoemission spectrum with a sum of many Gaussians with the width (full widths at half maxima) of 80 meV. Examples are shown in Fig.12. In Fig. 12(a), the angle-integrated EX photoemission spectrum measured at $\Delta t$=150 fs, shown by green curve, is analyzed using 21 Gaussians with 30-meV interval for the energy range between 1.8 to 2.4 eV. Thin solid black curves show the Gaussians, and open circles represent the deconvolved photoemission spectrum. At this time delay, when photoemissions from hot electrons in the CB still exist, only one peak below the CBM is resolved. We examined similar analysis using different numbers of Gaussians and/or different energy intervals. Although the peak height of each component of Gaussians changes slightly, the essential feature of resolved spectra does not change.

In Fig.12 (b), result of similar analysis of the data measured at $\Delta t$=217 fs is displayed. In this case, we used 81 Gaussians for the energy range from 1.8 to 2.4 eV (the energy interval in this case is 0.01 eV). For any spectra measured at $\Delta t$ ranging from 0 to 1.5 ps, only one peak with a significant asymmetry is resolved. Two-peak structures which show the maxima at energies of 1S- and 2S-exciton levels could not be resolved. Therefore, even in the early stage of exciton relaxation, the photoemission spectra show a single peak, and the peak energy and spectral shape are dependent on $\Delta t$.

**Appendix C: Exciton formation dynamics under 2.50-eV excitation**

We carried out similar time- and angle-resolved photoemission measurements by tuning $h\upsilon_{pump}$ to 2.50 eV in order to examine further the characteristics of exciton-formation dynamics in this crystal. Upon excitation at this photon energy, interband transitions from top-most valence band to the CB2 are not possible energetically. An estimation using the band structure in Fig. 3(c) gives the highest-energy state reached by the optical transition from the highest valence band is the state with excess energy of 0.25 eV above the CBM. However, transitions from the two top-most valence bands ($\Gamma_7^+$ and $\Gamma_8^+$) to the CB1 is dipole forbidden. Characteristics of optical transitions at 2.50 eV are substantially different from those at 3.40 eV. Therefore, the results under 2.50-eV excitation may provide useful information to reveal incoherent exciton-formation dynamics in $Cu_2O$. It is to be noted that $h\upsilon_{pump}$=2.50 eV is just below the resonance energy of the dipole-allowed 1S exciton state of Blue series.

Figure 13(a) shows angle-integrated photoemission spectra obtained at Δt=0 fs and 133 fs under excitation with $h\upsilon_{pump}$= 2.50 eV. At Δt=0 fs, a broad spectrum ranging from 2.4 to down to 1.2 eV is detected. This component is short-lived; it disappears at Δt=133 fs, as seen in Fig. 13(a). The solid green curve in Fig. 14(a) shows temporal changes in photoemission intensity at $1.55\pm0.04$ eV, which monitors exclusively the short-lived component. The response is identical to the CCT of pump-probe pulses shown by thin solid curve. Based on the results, we attribute the short-lived spectrum to a coherent two-photon photoionization by pump and probe pulses.

At Δt=133 fs, the EX photoemission peak is formed below the CBM, and it grows progressively, and the growth is associated with a peak-energy shift towards the low-energy side, as shown in Fig. 13(b). Figure 14 summarizes temporal evolution of the angle- and energy-integrated intensity (total intensity [58]) and of the peak energy of EX photoemission. Similar to the case of 3.40-eV excitation shown in Fig. 9, the total intensity grows to the maximum around Δt=1.5 ps, and the peak energy at Δt=1.5 ps converges to 2.04 eV, which is identical to the peak energy under 3.40-eV excitation. It is worthy to mention that we observe no signatures of photoemission around the energy of 1S exciton state of blue series.

As mentioned above, transitions from the two top-most valence bands ($\Gamma_7^+$ and $\Gamma_8^+$) to the first CB is dipole forbidden under 2.50-eV excitation. In fact, any signatures of hot electrons in the CB could not be detected as shown in Fig.13. Therefore, the density of e-h plasma, if any, is substantially lower than the case of $h\upsilon_{pump}$=3.40 eV. As seen in Fig.14 (b), the initial peak energy is 2.08 eV under 2.50-eV excitation, which is slightly lower than that (2.10 eV) under 3.40-eV excitation. The difference may be due to weaker effect of plasma screening to the e-h interaction under 2.50-eV excitation. Despite a significant difference in the density of e-h plasma initially generated by pump light, the relaxation process of EX photoemission under 2.50-eV excitation is very similar to that under 3.40-eV excitation displayed in Figs.4 and 9, indicating that the same process is responsible for the temporal evolution in the time frame of 1.5 ps for both cases.

A recent study on the optical property of Cu$_2$O has shown that the $\Gamma_3^-$-phonon-assisted transitions into the yellow- and green-1S exciton states contribute strongly to the optical transitions above 2.2 eV [66]. Therefore, it is likely that the phonon-assisted transitions result in the direct formation of excitons with finite $\vec{K}$ under 2.50-eV excitation, leading to a fast exciton formation. In fact, previous study on time-resolved Lyman spectroscopy under 2.33-eV excitation reported ultrafast formation of excitons within ~150 fs [67]. In the present measurements, the temporal width of 2.50-eV pump pulse is 70 fs. Therefore, excitons with finite $\vec{K}$ may be formed within the pump-pulse duration. Nevertheless, slow changes in the total intensity and in the peak kinetic energy of the EX photoemission in the time frame of 1.5 ps are evident in Figs. 13 and 14, showing important roles of cooling process of center-of-mass motion of excitons in incoherent exciton-formation dynamics in Cu$_2$O. We attempted to quantify the relaxation process of EX photoemission by introducing a relaxation time $\tau_p$. The blue curve in Fig. 14 (b) is a fit to the peak energy by an exponential-decay function $E_p(t) = \Delta_E \exp(-t/\tau_p) + 2.04$ with $\tau_p$ = 0.50 ps; $\Delta_E$=0.054 eV is assumed. It describes satisfactorily the temporal evolution of peak energy of EX photoemission. Similarly, the blue curve in Fig. 14 (a), which can serve as a guide to the eye, is a fit of the function $I_0\{1-\exp(-t/\tau_p)\}$ to the total intensity of the EX photoemission with $\tau_p$=0.51 ps. Although the analysis is phenomenological, we can conclude that the cooling process of center-of-mass motion of excitons is characterized by a time constant of about 0.50 ps.

*Present address: Institute for Materials Research, Tohoku University, 2-1-1 Katahira, Aoba-ku, Sendai 980-8577, Japan.

# tanimura@sanken.osaka-u.ac.jp

Figure captions

Fig. 1 A schematic diagram of the photoionization process of excitons. The kinetic energy of an electron photoionized from an exciton is governed by the recoil effects of the conjugate hole, which is determined by the valence-band dispersion. For symbols used in the figure see the text.

Fig.2 A 170 Å x 230 Å STM image of $Cu_2O$(111) surface, acquired with a negative tip-bias voltage (-1.5 V) and tunneling current of 0.03 nA. An example of randomly distributed protrusions is highlighted by a white circle labeled P1, and typical crystallographic directions are illustrated by arrows.

Fig.3 (a) and (b) Photoemission intensity maps measured at Δt=17 fs and Δt=500 fs for p-polarized 3.40-eV pump pulses at 90 K; the photoemission intensities are plotted as a function of initial-state energy ε and the $k_\parallel$. The ε is referenced with the VBM. The solid white curves show the dispersion of the first, second and third conduction bands along the Γ-X direction with the CBM energy of 2.17 eV. Rectangles labeled A to G indicate representative positions with different ε and wave vectors in the conduction band. In (a) and (b), the color scale indicates the photoemission intensity. (c) A part of the band structure of $Cu_2O$. The band dispersions calculated in Ref. 47 were plotted with the experimental values of the band gap (2.17 eV) and the minimum of the second CB (2.62 eV). Violet arrows show estimated optical transitions induced by3.40-eV photons to the highest-energy position in the conduction band. (d) Experimental geometry of photoemission measurements, and the relation between the surface Brillouin zone and bulk Brillouin zone for $Cu_2O$ with (111) surface under this geometry. Positive θ corresponds to the right half from $\overline{\Gamma}$ to $\overline{M}'$.

Fig.4 Temporal evolution of angle-integrated photoemission spectra under 3.40-eV excitation. In (a), the changes above 2.2 eV are highlighted on an expanded vertical scale, while (b) displays the changes in EX-photoemission spectra on an expanded energy scale from 1.8 to 2.5 eV. In (b), the curves are offset for clarity by a constant value, and the base line is shown by red line for each curve. The typical time delays of spectra are shown by number in unit of ps.

Fig.5 (a) Two-dimensional photoemission intensity map measured at $\Delta t$=1.5 ps; the photoemission intensities are plotted as a function of kinetic energy and $k_{//}$. The solid white curve shows the calculated kinetic energy using Eq. (1) with the valence-band dispersion along the Γ-X direction, while the broken curve shows the dispersion of the first conduction band along the Γ-X direction (from Ref.49). (b) The energy-integrated EX-photoemission intensity plotted as a function of $k_{//}$. Blue, black, and green curves show the intensity distributions calculated using Eq.(5) with $a_{ex}$ =4.5 Å, 5.3 Å, and 6.5 Å.

Fig.6 (a) The top-most valence band dispersion along the Γ-R direction (from Ref.49). The energy at the Γ point is set to zero. (b) The green curve shows the normal photoemission spectrum measured at $\Delta t$= 1.5 ps after 3.40-eV photoexcitation. The intensities for $-1° \leq \theta \leq +1°$ are integrated. The red curve shows the calculated normal photoemission spectrum assuming the hydrogenlike 1S-state wavefunction with $a_{ex}$ =5.3 Å (for details see the text). The solid black curve shows the result of convolution of the calculated normal photoemission spectrum with respect to the energy resolution (80 meV) in the measurement. The peak kinetic energy (1.37 eV) of the calculated normal photoemission spectrum is adjusted in such a way that the convoluted spectrum gives the peak kinetic energy same as the experimental result. The intensities are normalized with respect to the peak intensities.

Fig.7 (a) Temporal changes in the photoemission intensities at the regions A and B in Fig.2. The excess energy is 0.95 eV above the CBM, and $k_{||}$=0.18 Å$^{-1}$ (0 Å$^{-1}$) at A (B). The solid black curve is a result of rate-equation analysis with decay time of 38 fs. (b) Temporal changes in the photoemission intensities at the regions C, D, E in Fig.2. The excess energy is 0.63 eV above the CBM, and $k_{\parallel}$=0.17, 0.09 and 0 Å$^{-1}$ at C, D and E. (c) Temporal changes in the photoemission intensities at the regions F and G in Fig.2. The excess energy is 0.26 eV above the CBM, and $k_{\parallel}$=0.14 Å$^{-1}$ (0 Å$^{-1}$) at E (F). (d) The angle-integrated photoemission intensities, plotted on a semi-logarithmic scale, at the states with typical excess energies above the CBM. The broken curve shows the cross-correlation trace between pump and probe pulses, and the solid black curve for the data at $E_{ex}$=1.15 eV is the result of rate-equation analysis. The solid black line for the data at $E_{ex}$=0.13 eV shows an exponential decay with a time constant of 87 fs.

Fig.8 (a) Population decay times of hot electrons in the conduction band as a function of the excess energy above the CBM. The green dots are the decay times determined by semi-logarithmic fit of the temporal changes in photoemission intensity, while the open circles are those estimated by a rate-equation analysis assuming a single-exponential decay of the intensity. The solid curve shows the result of Eq. (6) (see the text). (b) The dispersions of conduction bands (solid black curves) along the Γ-X direction calculated in Ref. 49. The broken red curve is a fit of Eq. (7) to the CB1.

Fig.9 (a) Temporal changes in the angle- and energy-integrated photoemission intensities for the hot electrons in the CB (green curve) and the EX photoemission (blue curve). The thin solid curve is a fit to the data of hot electrons of a rate-equation analysis, while the broken curve shows the density of hot electrons generated by a pump pulse, which was estimated by integrating a pump pulse without any decay terms. The constant level corresponds to the excitation density of 2.5x10$^7$ cm$^{-3}$. The red curve show the difference between broken and solid black curves, which shows the growth of a product generated by electron-hole combination. (b) The peak energy of EX photoemission as a function of time delay under 3.40-eV excitation (blue curve). The energies of 1s and 2p states of the yellow-series excitons and that of 2s state of the green series are shown for comparison. (c) The width of momentum spread of EX photoemission as a function of time delay. The solid line shows the value of quasi-steady state reached at Δt=1.5 ps

Fig.10 (a)~(c): Two-dimensional photoemission intensity map measured at Δt=120 fs; 205 fs, and 320 fs. The photoemission intensities specified by a color scale are plotted as a function of $E_K$ and $k_\parallel$. The solid white curves in (a), (b) and (c) show the exciton dispersion curve with the mass of center-of-mass motion of $1.57 m_0$ (see the text). (d)~(f): Momentum-resolved photoemission spectra at $k_\parallel$=0, red, and $k_\parallel$=0.10 Å$^{-1}$, blue, for the results in (a)~(c). (g): Energy-integrated photoemission intensities as a function of $k_\parallel$ at Δt=120 fs, 205 fs, 320 fs, and 1.5 ps. Energy ranges for integration are $E_P(0) \pm 0.1$ eV, where $E_P(0)$ is the peak kinetic energy of the momentum-resolved spectra at $k_\parallel$=0 Å$^{-1}$.

Fig.11 The photoemission spectrum probed by p-polarized 6.00-eV laser light for Cu$_2$O:(111)-(1x1) at 90 K. The electron kinetic energy is referenced with respect to the vacuum level of analyzer. The low-energy cutoff is at $E_k$=0.83 eV. The high-energy cutoff ($E_k$=1.05±0.01 eV) was determined by linear extrapolations as shown by the data presented on an expanded scale (green curve).

Fig. 12 Results of deconvolution analysis for the EX photoemission spectra measured at Δt=150 fs, (a), and Δt=217 fs, (b). In (a) and (b), solid green curves show the angle-integrated EX photoemission spectra. In (a), the observed photoemission spectrum was fitted with a sum of 21 Gaussians, each of which has a constant width (full widths at half maxima) of 80 meV, in the energy range between 1.8 to 2.4 eV. The Gaussians with best-fit amplitude (shown by red circles) are shown by thin solid curves, and the sum of the intensities is shown by solid red curve. In (b), 81 Gaussians with 80-meV widths were used in the analysis, and their best-fit amplitudes are plotted by red circles; the sum of intensities is shown by the solid red curve.

Fig.13 Temporal evolution of angle-integrated photoemission spectra under 2.50-eV excitation. In (a), the spectra measured at Δt=0 fs and Δt=133 fs are compared. The energies of the CBM and the 1s exciton in the blue series are indicated by arrows. In (b), the changes in EX-photoemission spectra are displayed on an expanded energy scale from 1.75 to 2.42 eV. The curves are offset for clarity by a constant value, and the base line is shown by red line for each curve with Δt (in unit of ps) when the spectra were measured. The energies of 2p state of the yellow series and that of the CBM are shown for comparison.

Fig.14 (a) Temporal change in the angle- and energy-integrated EX photoemission intensities (open circle) and in the energy-resolved intensity at ε=1.4±0.04 eV (green curve) under 2.50-eV excitation. The solid black curve is the Gaussian-fitted CCT trace of pump- and probe pulses. The blue curve, is a fit of the function $I_0\{1-\exp(-t/\tau_p)\}$ to the total intensity of the EX photoemission with $\tau_p$=0.51 ps. (b) The peak energy of EX photoemission as a function of time delay under 2.50-eV excitation (open circles). The solid blue curve shows a fit by an exponential-decay function with a single time constant of 0.50 ps (see the text). The energies of 1s state of the yellow-series excitons and of 2s state of the green series are shown for comparison.

Fig.1 H. Tanimura et al.

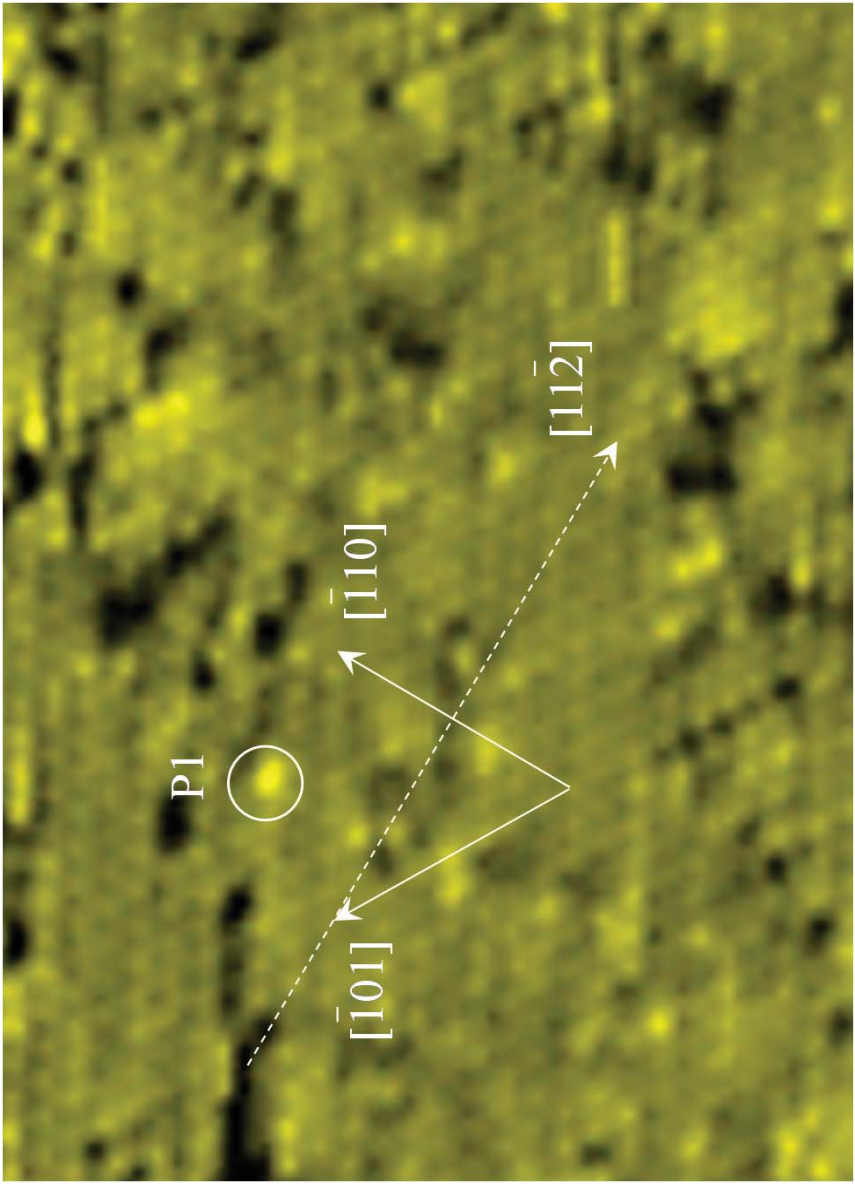

Fig.2 H. Tanimura et al.

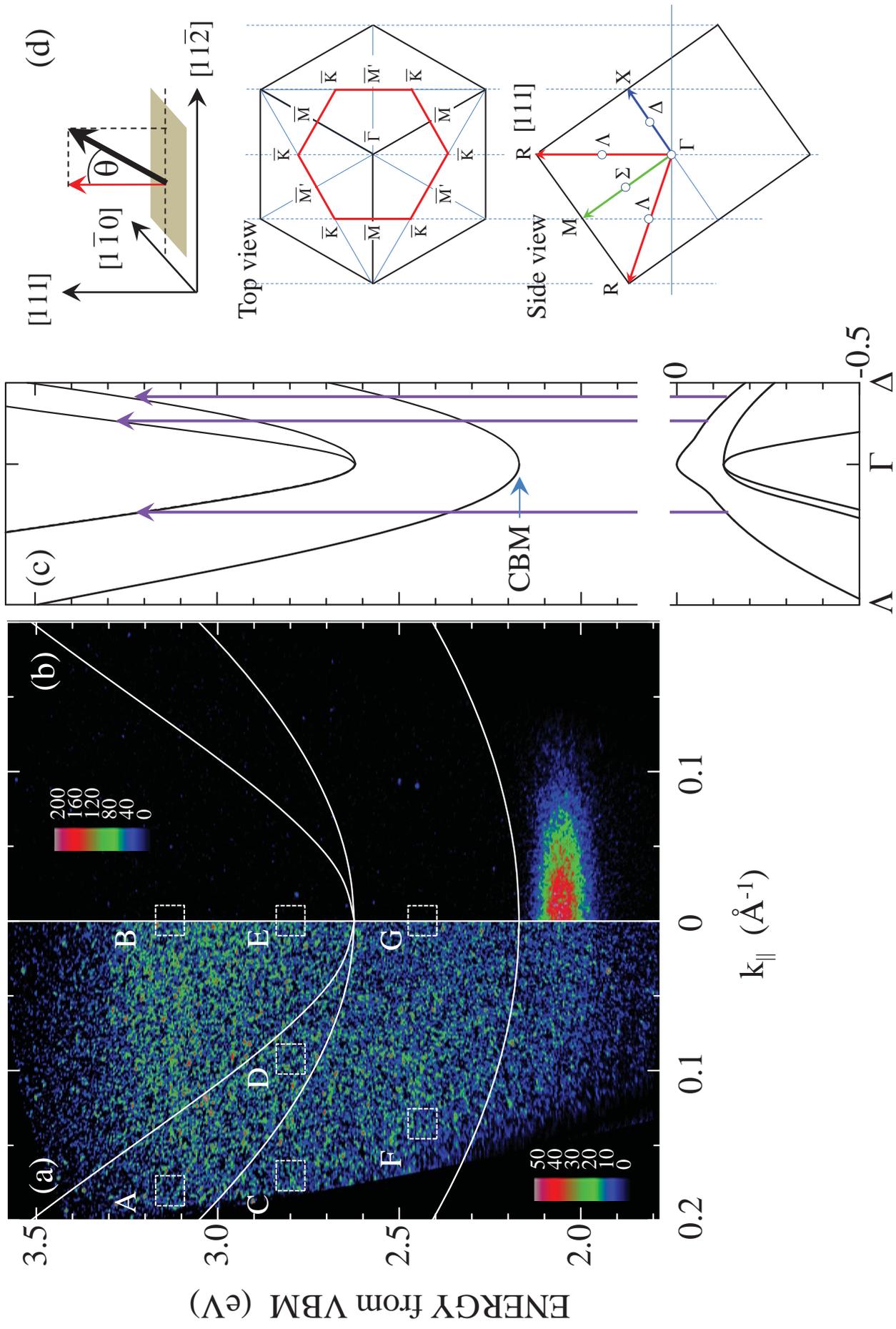

Fig.3 H. Tanimura et al.

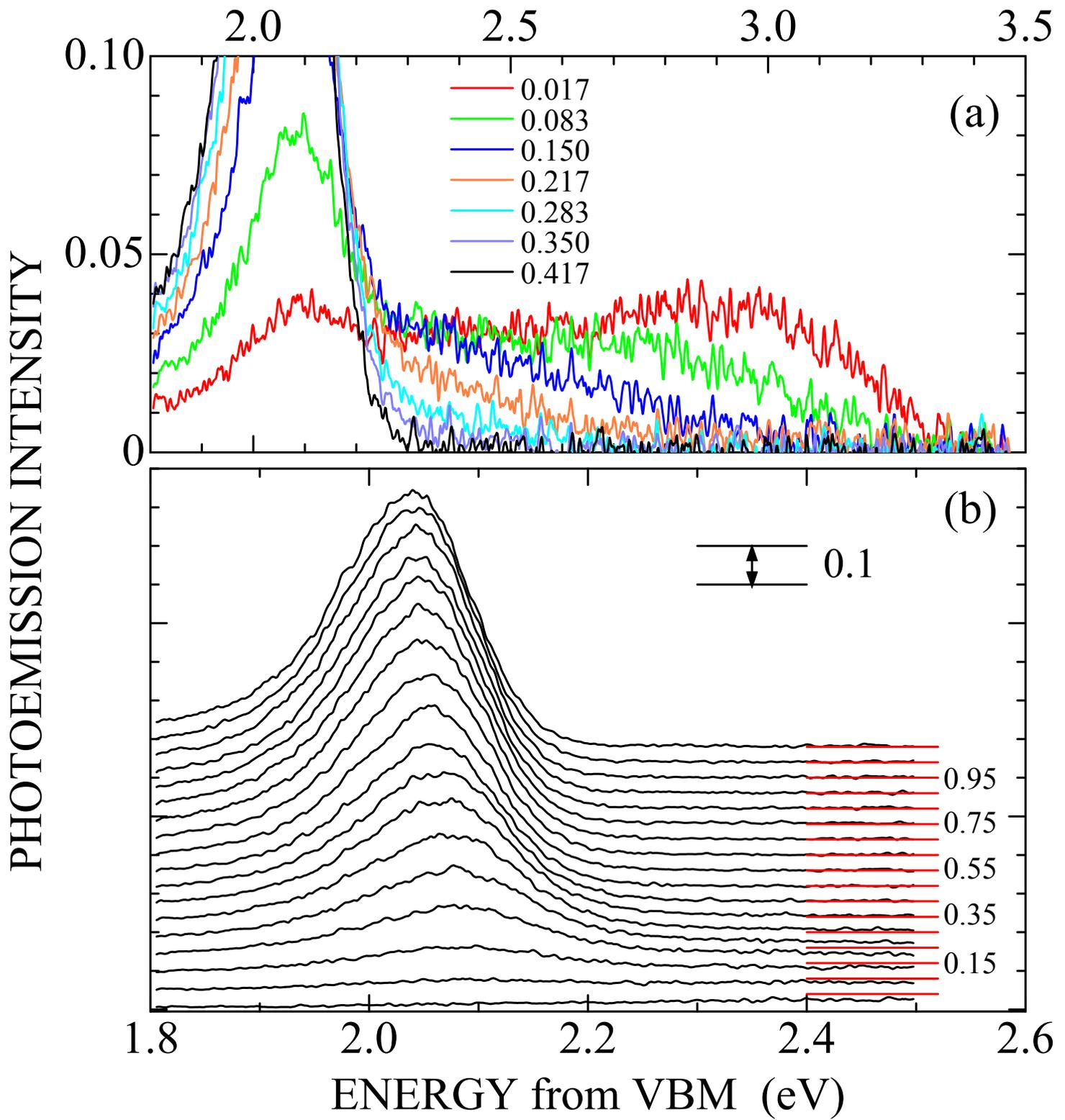

Fig.4 H. Tanimura, et al.

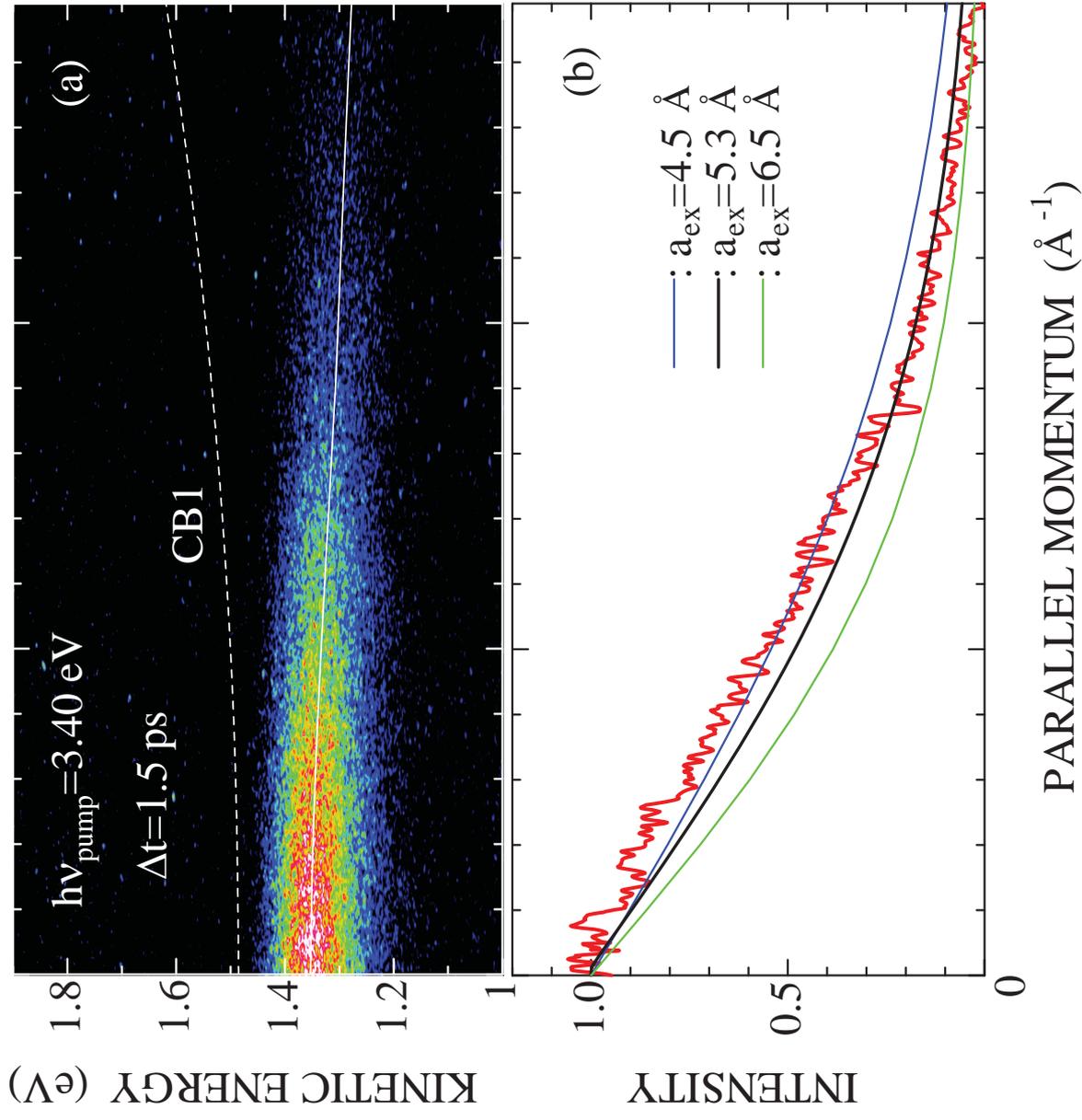

Fig.5 H. Tanimura et al.

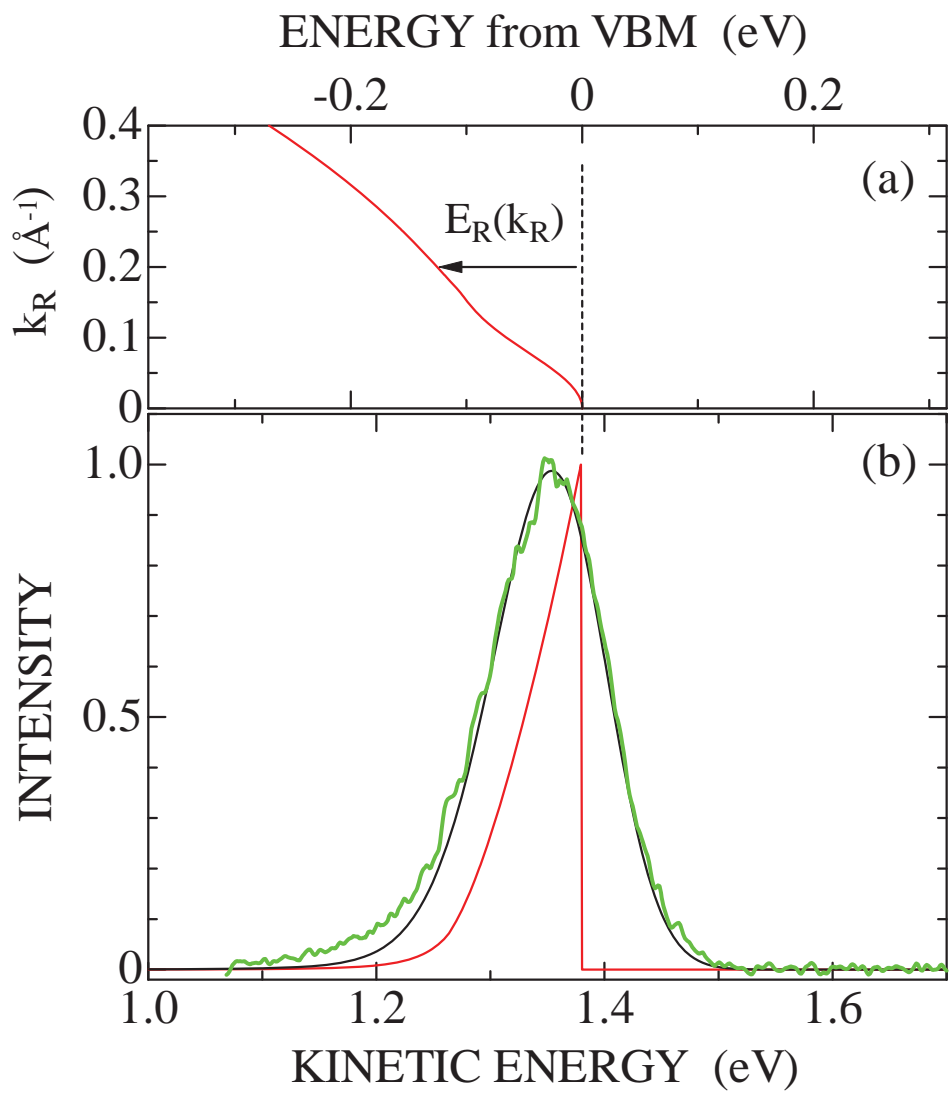

Fig.6 H. Tanimura et al.

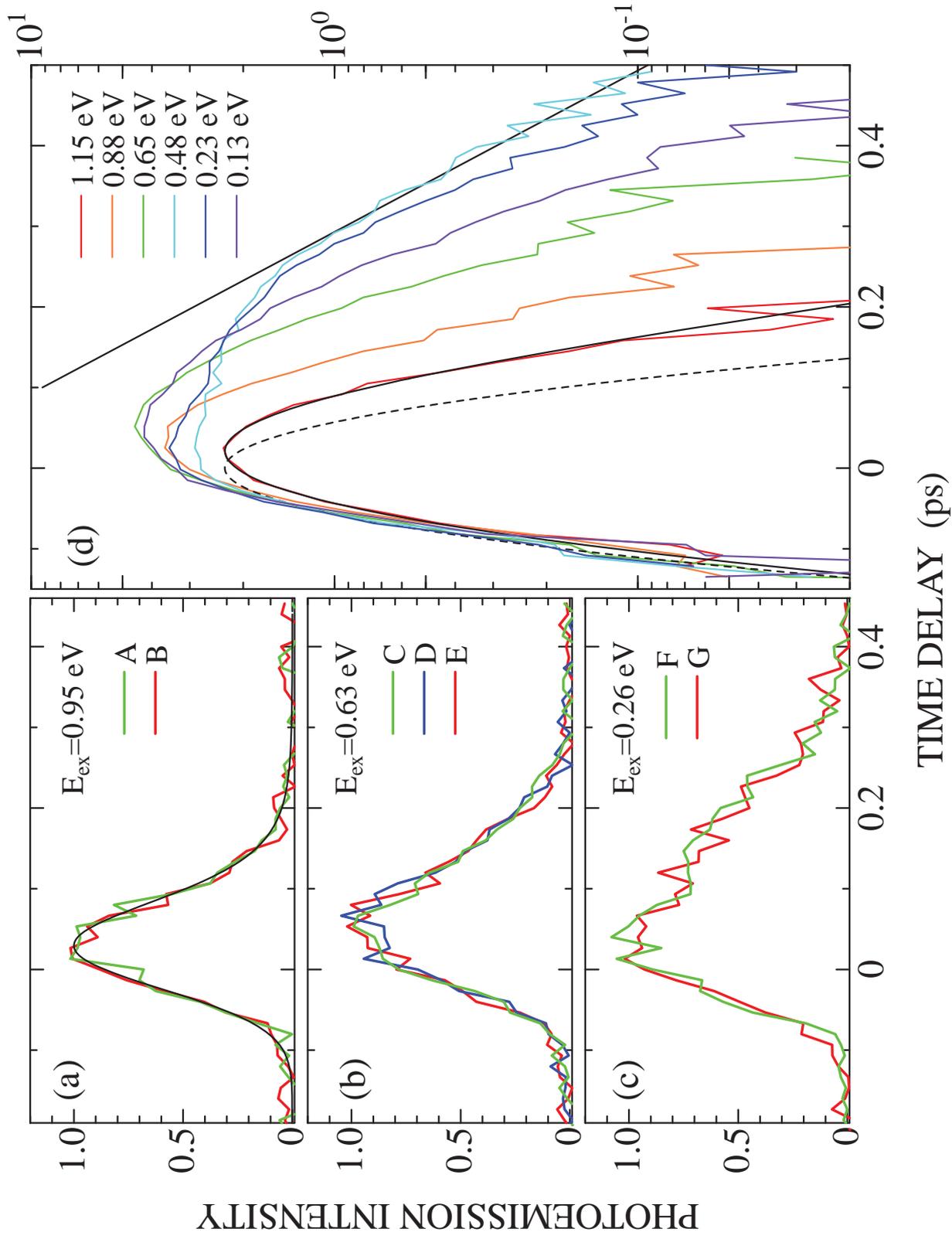

Fig.7 H. Tanimura et al.

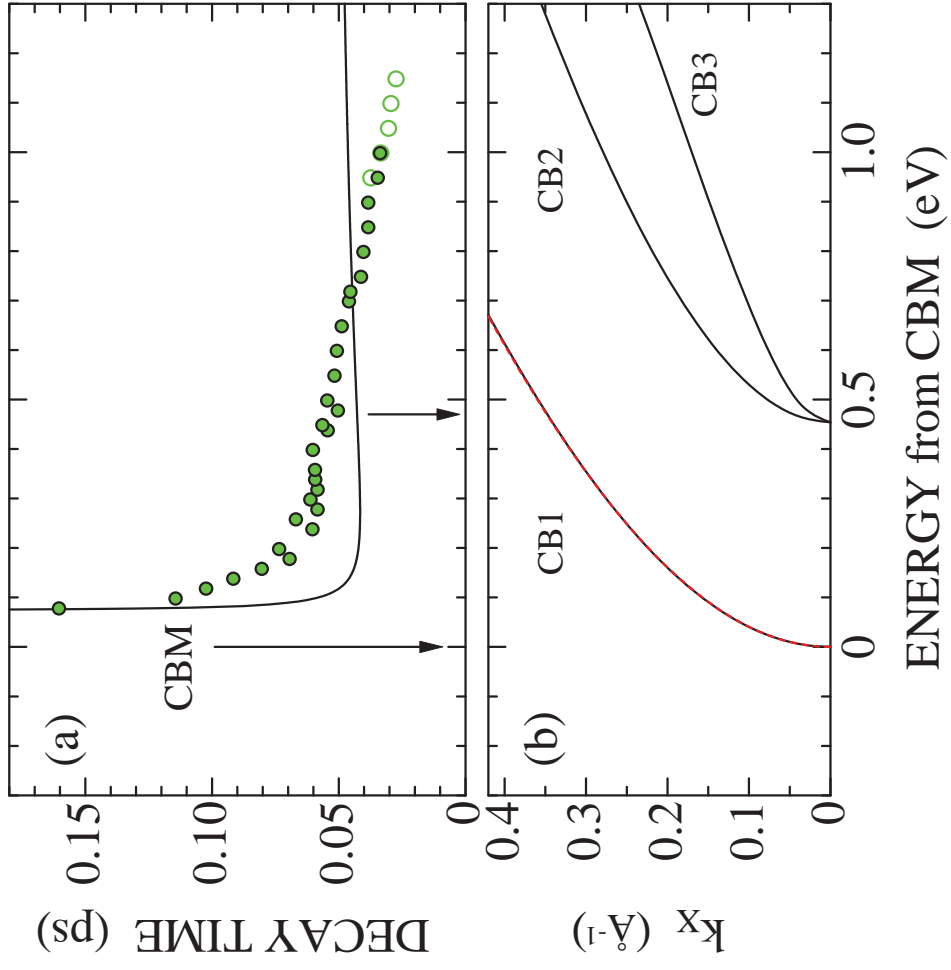

Fig.8  H. Tanimura et al.

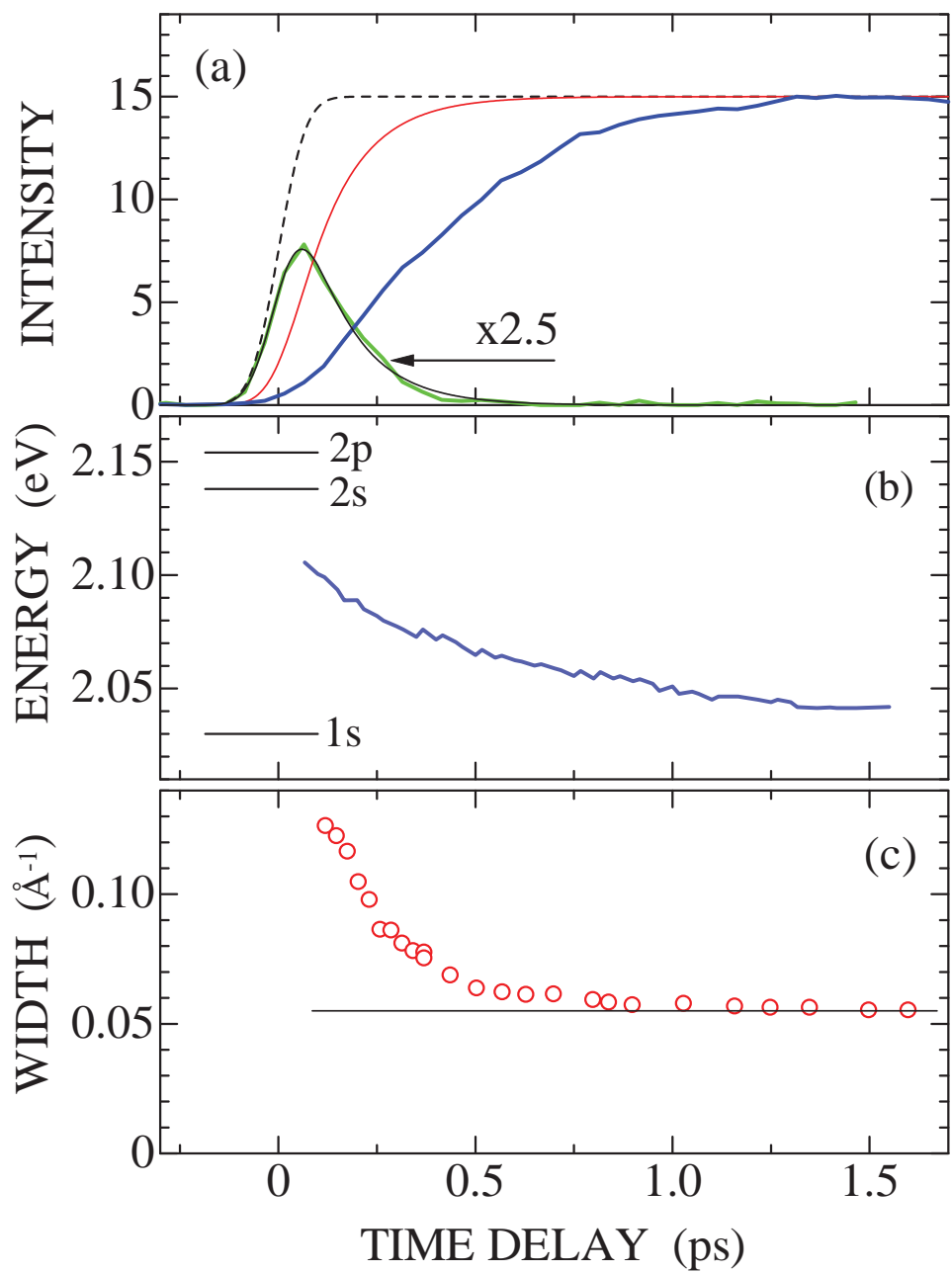

Fig.9 H. Tanimura et al.

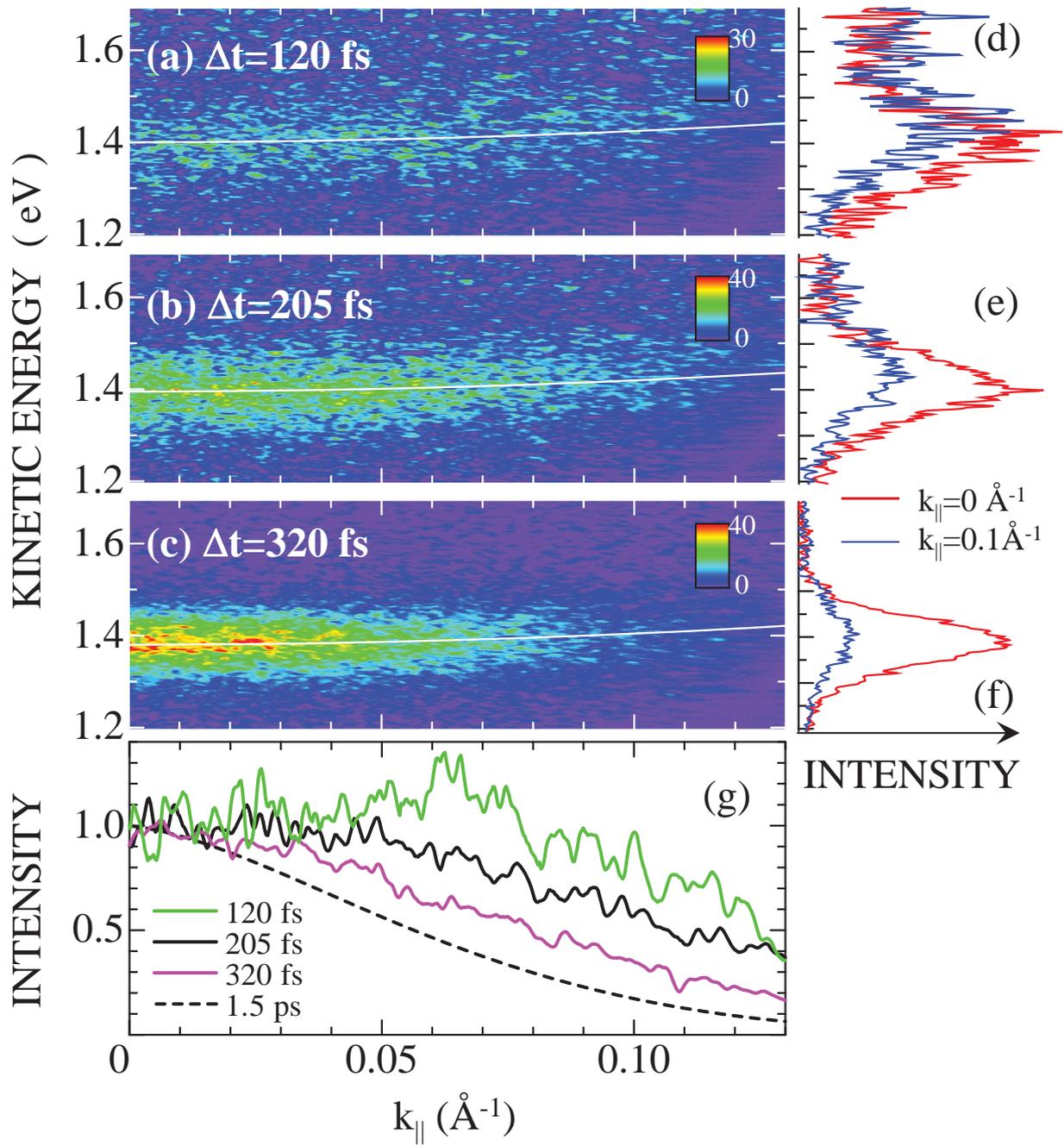

Fig.10 H. Tanimura et al.

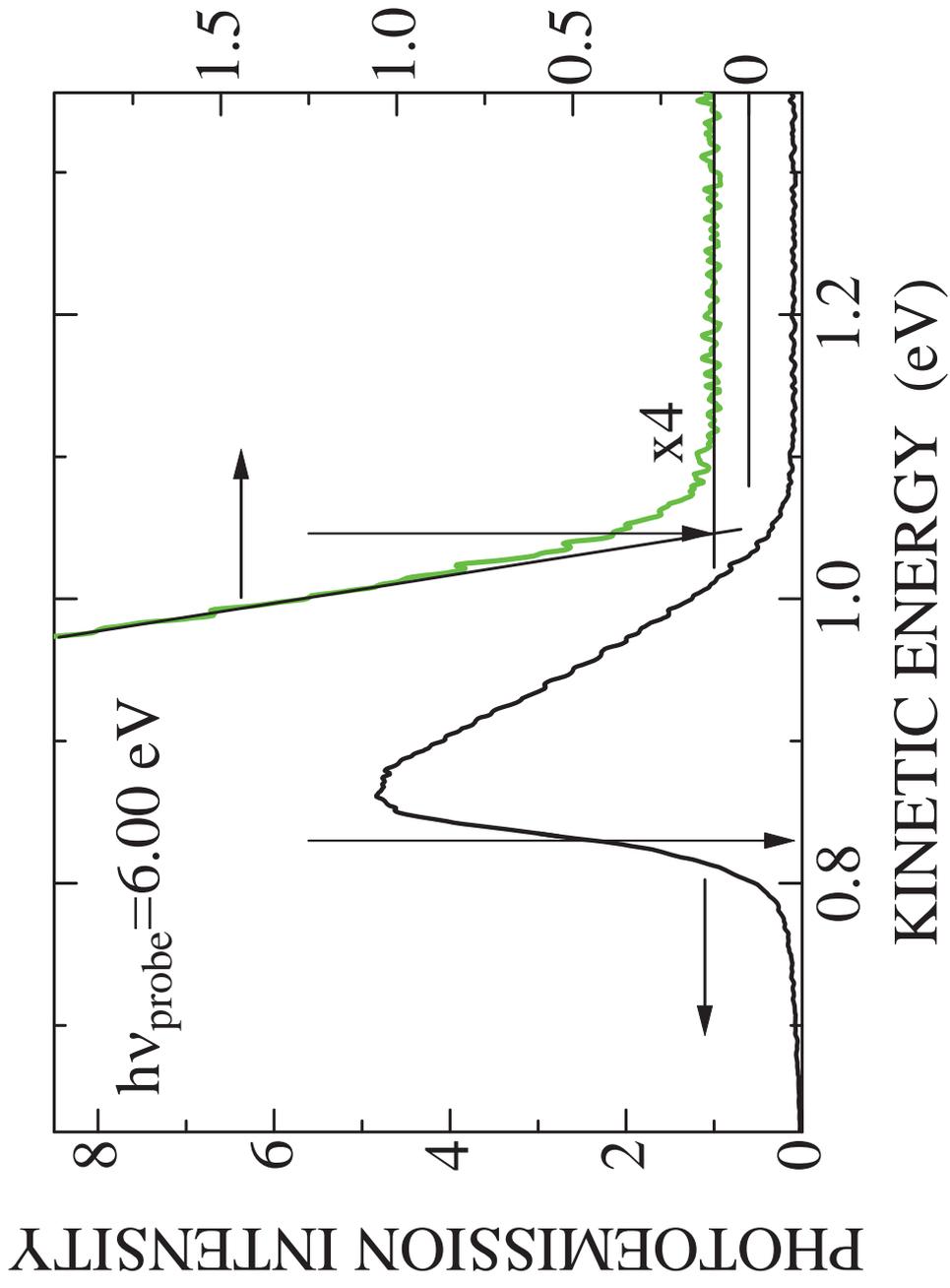

Fig.11 H. Tanimura et al.

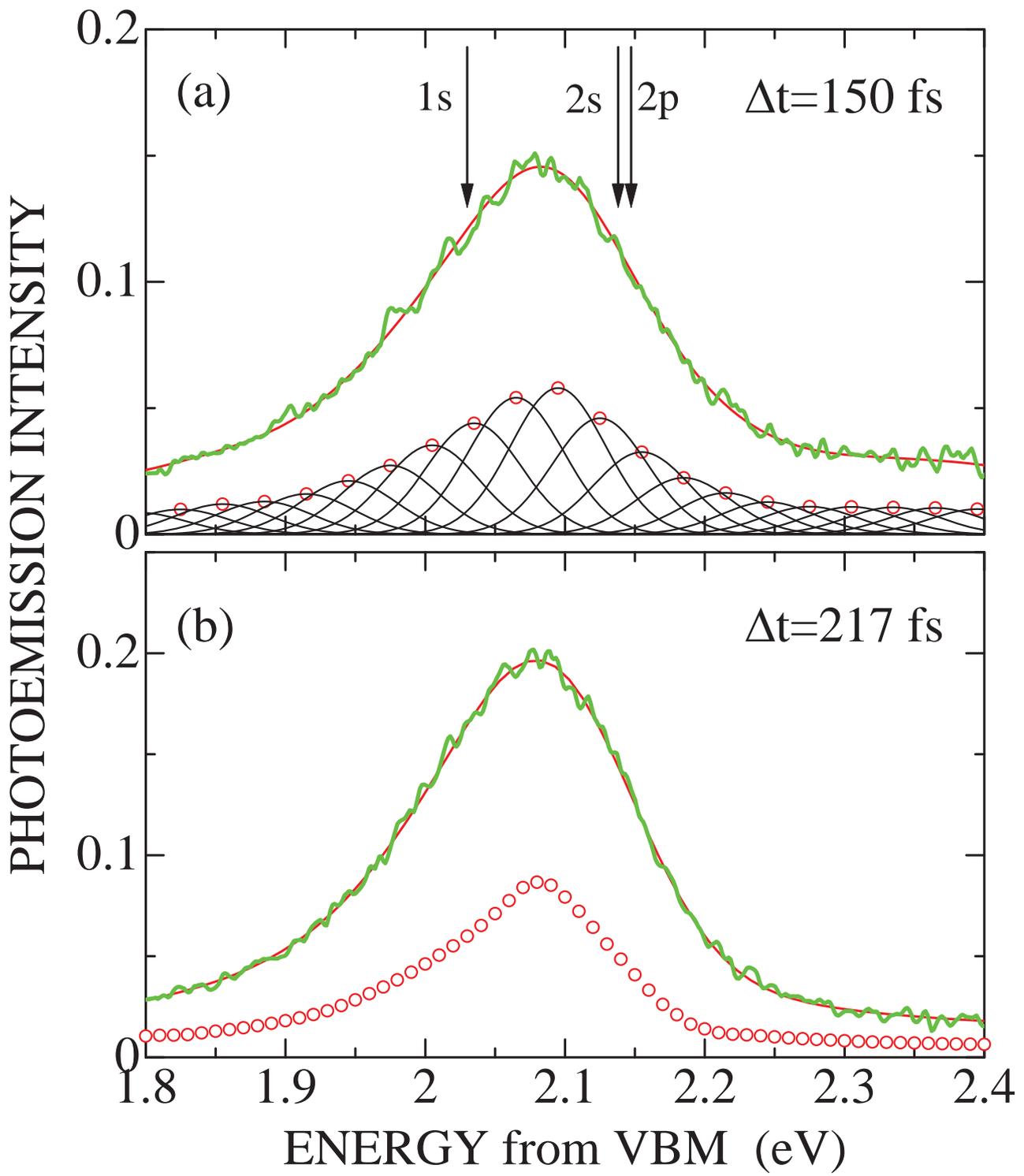

Fig.12 H. Tanimura et al.

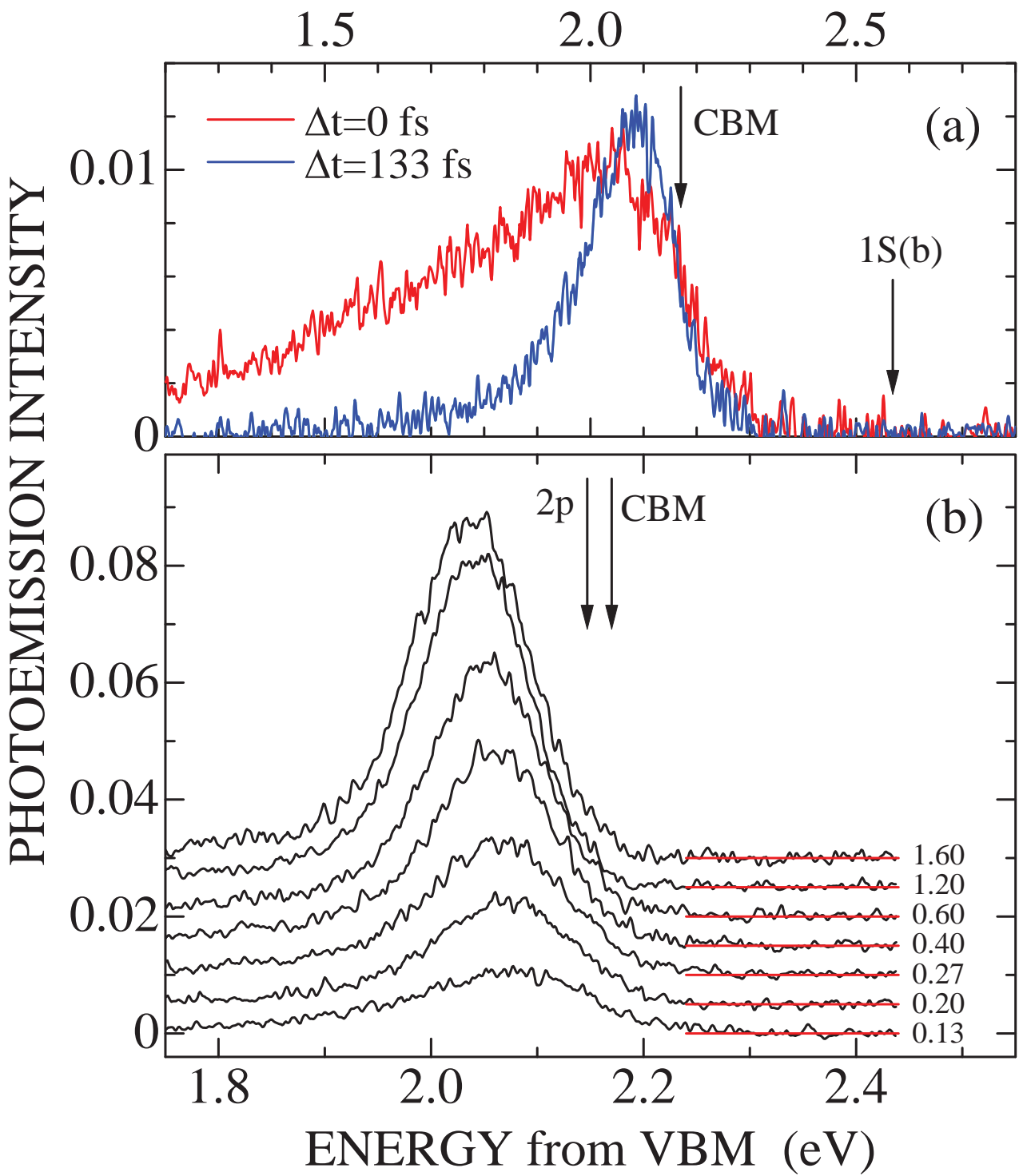

Fig.13 H. Tanimura. et al.

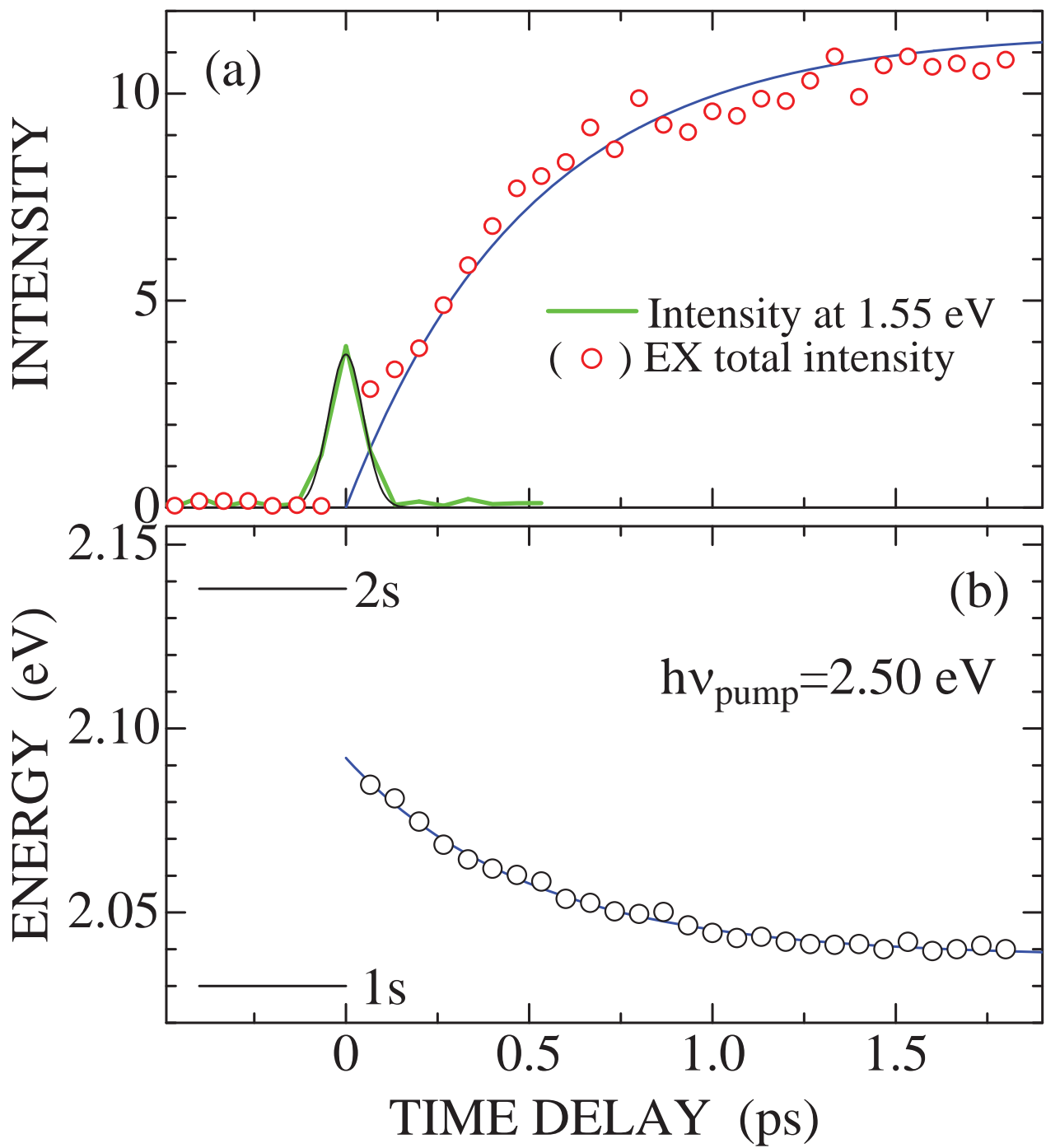

Fig.14 H. Tanimura et al.